\documentclass[aps,twocolumn,floatfix,superscriptaddress,preprintnumbers,showpacs,showkeys]{revtex4}
\usepackage{amssymb}
\usepackage{amsmath}
\usepackage{amsfonts}
\usepackage{epsfig}
\usepackage{graphicx}
\usepackage{tabularx}
\newcommand{\seq}{\begin{subequations}}
\newcommand{\sen}{\end{subequations}}
\newcommand{\eq}{\begin{eqnarray}}
\newcommand{\en}{\end{eqnarray}}

\def\shiftdown#1{#1\llap{\lower.04ex\hbox{#1}}}

\newcommand{\ra}{\rangle}
\newcommand{\la}{\langle}

\newcommand{\bfD}{{\bf \Delta}_{\perp}}
\newcommand{\bfq}{{\bf q}_{\perp}}

\newcommand{\bfk}{{\bf k}_{\perp}}

\newcommand{\bfb}{{\bf b}_{\perp}}

\begin{document}

\title{Nucleon parton distributions in a light-front quark model} 

\author{Thomas Gutsche}
\affiliation{Institut f\"ur Theoretische Physik,
Kepler Center for Astro and Particle Physics,  
Universit\"at T\"ubingen, 
Auf der Morgenstelle 14, D-72076 T\"ubingen, Germany}
\author{Valery E. Lyubovitskij} 
\affiliation{Institut f\"ur Theoretische Physik,
Kepler Center for Astro and Particle Physics,  
Universit\"at T\"ubingen, 
Auf der Morgenstelle 14, D-72076 T\"ubingen, Germany}
\affiliation{Department of Physics, Tomsk State University,  
634050 Tomsk, Russia} 
\affiliation{Laboratory of Particle Physics, 
Mathematical Physics Department, 
Tomsk Polytechnic University, 634050 Tomsk, Russia} 
\affiliation{Departamento de F\'\i sica y Centro Cient\'\i fico 
Tecnol\'ogico de Valpara\'\i so (CCTVal), Universidad T\'ecnica
Federico Santa Mar\'\i a, Casilla 110-V, Valpara\'\i so, Chile}
\author{Ivan Schmidt}
\affiliation{Departamento de F\'\i sica y Centro Cient\'\i fico 
Tecnol\'ogico de Valpara\'\i so (CCTVal), Universidad T\'ecnica
Federico Santa Mar\'\i a, Casilla 110-V, Valpara\'\i so, Chile}

\date{\today}

\begin{abstract}

Continuing our analysis of parton distributions in the nucleon, 
we extend our light-front quark model in order to obtain both 
the helicity-independent and the helicity-dependent parton distributions, 
analytically matching the results of global fits at the initial scale 
$\mu \sim 1$ GeV; they also contain the correct
Dokshitzer-Gribov-Lipatov-Altarelli-Parisi evolution.  
We also calculate the transverse parton, Wigner and Husimi distributions
from a unified point of view, using our light-front wave functions and
expressing them in terms of the parton distributions $q_v(x)$ and
$\delta q_v(x)$. Our results are very relevant for the current
and future program of the COMPASS experiment at SPS (CERN).

\end{abstract}



\maketitle

\section{Introduction}

In Refs.~\cite{Gutsche:2013zia,Gutsche:2014yea} we
proposed phenomenological light-front wave functions (LFWFs)
for the nucleon, which produce a description of electromagnetic
form factors of nucleons consistent with data and with
the correct power behavior at higher 
scales~\cite{Brodsky:1973kr,Matveev:1973kr}. 
The difference in the two papers~\cite{Gutsche:2013zia,Gutsche:2014yea} 
concerns the modeling of the $x$-dependence, 
which has an impact on the scaling behavior of nucleon parton distributions. 
In the first case~\cite{Gutsche:2013zia} the nucleon parton distributions 
have the correct $x$ behavior at large scales, while
at the initial scale $\mu \sim 1$ GeV they were different from
the results of the world data analysis. 
In the second paper~\cite{Gutsche:2014yea}, 
we improved the $x$-dependence of the LFWFs in a such way  
that the modified LFWFs produced the correct helicity-independent 
parton distributions at the starting point for the 
Dokshitzer-Gribov-Lipatov-Altarelli-Parisi (DGLAP) 
evolution~\cite{DGLAP1,DGLAP2,DGLAP3,DGLAP4}.  
In the latter case~\cite{Gutsche:2014yea} 
we also had some freedom in setting up the LFWFs, 
because we did not consider helicity-dependent parton distributions.  
A similar application of the pion LFWFs, resulting in good 
agreement with data and in the correct scaling of form factors and parton 
distributions, has been made in Ref.~\cite{Gutsche:2014zua}.  
Note that the problem of constructing of the nucleon LFWFs 
was extensively studied in the literature starting from 
a pioneer paper by Brodsky et al.~\cite{Brodsky:1982nx}   
and continuing by progress done by many groups in derivation 
of the LFWFs and its applications to nucleon phenomenology 
(see e.g. Refs.\cite{LFQCD1}-\cite{Chakrabarti:2016yuw}). 

In the present manuscript we derive the nucleon LFWFs where now
the $x$-dependence is encoded by knowledge of the helicity-independent 
$q_v(x)$ and helicity-dependent $\delta q_v(x)$ valence parton distributions. 
The main advantage of our approach is that the derived LFWF does not depend
on phenomenological parameters like masses of quark/diquark,
which are not directly related to QCD. Restricting to zero
current quark masses we obtain a reasonable description of
data on nucleon form factors.  
The paper is organized as follows. 
In Sect.~II we construct the nucleon LFWFs, which will be used for 
the calculation of parton distributions and form factors using 
the presentations of these quantities in terms of the LFWFs. 
In Sect.~III we collect the well-known decompositions 
of the nucleon Dirac and Pauli form factors, parton distributions 
(including longitudinal, transverse, Wigner and Husimi distributions) 
in terms of the LFWFs. In Sect.~IV we present our numerical results 
and discussion. Finally, Sect.~V contains our summary and conclusions. 
We have collected some technical material on the Wigner and Husimi
parton distributions in the appendix. 

\section{Nucleon light-front wave functions}
\label{Sec_Nucleon_LFWF}

For simplicity we consider the quark-scalar diquark model, where  
the generic ansatz for the massless
LFWFs at the initial scale $\mu_0 = 1$ GeV 
reads~\cite{Gutsche:2013zia,Gutsche:2014yea} 
\eq\label{LFWF_in}
\psi_{+q}^+(x,\bfk) &=& \varphi_q^{(1)}(x,\bfk)
\,, \nonumber\\
\psi_{-q}^+(x,\bfk) &=& - \frac{k^1 + ik^2}{M_N} 
\, \varphi_q^{(2)}(x,\bfk)  \,, \nonumber\\
& &\\
\psi_{+q}^-(x,\bfk) &=& \frac{k^1 - ik^2}{M_N}  
\, \varphi_q^{(2)}(x,\bfk)
\,, \nonumber\\
\psi_{-q}^-(x,\bfk) &=& \varphi_q^{(1)}(x,\bfk)
\,, \nonumber
\en
where $\varphi_q^{(1)}$ and $\varphi_q^{(2)}$ are
the LFWFs: 
\eq 
\varphi_q^{(1)}(x,\bfk) &=& \frac{4 \pi}{M_N} \, 
\sqrt{\frac{q_v(x) + \delta q_v(x)}{2}} \, \sqrt{D_q^{(1)}(x)} \nonumber\\ 
&\times& \exp\biggl[- \frac{\bfk^2}{2 M_N^2} D_q^{(1)}(x) \biggr] \,, 
\en 
\eq 
\varphi_q^{(2)}(x,\bfk) &=& \eta_q \, \frac{4 \pi}{M_N} \, 
\sqrt{\frac{q_v(x) - \delta q_v(x)}{2}} \, D_q^{(2)}(x) \nonumber\\ 
&\times& \exp\biggl[- \frac{\bfk^2}{2 M_N^2} D_q^{(2)}(x) \biggr] \,.  
\en 
Here $M_N$ is the nucleon mass, 
$q_v(x)$ and $\delta q_v(x)$ are the helicity-independent and 
helicity-dependent valence quark parton distributions 
(for these quantities the exact expressions 
from a world data analysis at the initial scale are understood), 
$D_q^{(1)}$ and $D_q^{(2)}$ are the longitudinal wave functions, 
connected to the electromagnetic form factors of the nucleon, $\eta_u = 1$ 
and $\eta_d = -1$. 
Therefore, in our  ansatz for the nucleon LFWFs 
only the functions $D_q^{(1)}(x)$ and $D_q^{(2)}(x)$ 
are free to be modeled. 
The parameterizations for $D_q^{(1)}(x)$ and $D_q^{(2)}(x)$ 
are not necessary the same and could be different, because these functions
parametrize the LFWFs with different helicities of quark and nucleon 
(see below). 
Note that the nucleon mass $M_N$ is chosen as the scale parameter. 
It is easy to see that the choice of a specific scale $\Lambda$ 
in Eq.~(\ref{LFWF_in}) is not important. 
A change of the scale $\Lambda \to \tilde\Lambda = \alpha \, \Lambda$ 
in~(\ref{LFWF_in}) can be compensated for by a rescaling of the longitudinal 
functions $D_q^{(1)}$ and $D_q^{(2)}$ as 
\eq 
D_q^{(i)} \to \tilde D_q^{(i)} = \alpha^2 \, D_q^{(i)} \,.
\en  
For simplicity we therefore choose a scale coinciding with the nucleon mass 
$\Lambda = M_N$. 
Our functions $\varphi_q^{(1)}$ and $\varphi_q^{(2)}$ are normalized as 
\eq\label{norm_cond1} 
\int \frac{d^2\bfk}{16 \pi^3} \, \Big[\varphi_q^{(1)}(x,\bfk)\Big]^2 
&=& \frac{q_v(x) + \delta q_v(x)}{2} \,, \nonumber\\
\int \frac{d^2\bfk}{16 \pi^3} \, \frac{\bfk^2}{M_N^2} \, 
\Big[\varphi_q^{(2)}(x,\bfk)\Big]^2 &=& 
\frac{q_v(x) - \delta q_v(x)}{2} 
\en 
and 
\eq\label{norm_cond2}  
\int\limits_0^1 dx 
\int \frac{d^2\bfk}{16 \pi^3} \, \Big[\varphi_q^{(1)}(x,\bfk)\Big]^2 &=& 
\frac{n_q + g_A^q}{2}  \,, \nonumber\\ 
\int\limits_0^1 dx 
\int \frac{d^2\bfk}{16 \pi^3} \, \frac{\bfk^2}{M_N^2} \, 
\Big[\varphi_q^{(2)}(x,\bfk)\Big]^2 &=& 
\frac{n_q - g_A^q}{2}  \,,
\en 
where $n_q$ is the number of $u$ or $d$ valence quarks in the proton and 
$g_A^q$ is the axial charge of a quark with flavor $q=u$ or $d$. 

The functions $\varphi_q^{(1)}$ and $\varphi_q^{(2)}$ 
are generalizations of the LFWFs
found by matching the electromagnetic form factors of the nucleon in
soft-wall AdS/QCD~\cite{Brodsky:2007hb}-\cite{Gutsche:2013vb} and
light-front QCD (see the detailed discussion 
in Refs.~\cite{Gutsche:2013zia,Gutsche:2014yea}).
In particular, as a result of the matching procedure the functions 
$\varphi_q^{(i)}(x,\bfk)$ have been deduced: 
\eq
\varphi_q^{{\rm AdS/QCD} (i)}(x,\bfk) &\sim&
\frac{4\pi}{\kappa} \,
\frac{\sqrt{\log(1/x)}}{1-x} \nonumber\\
&\times&\exp\biggl[- \frac{\bfk^2}{2\kappa^2}
\, \frac{\log(1/x)}{(1-x)^2}
\biggr] \,. 
\en
Note that the derived LFWF is not symmetric under the exchange
$x \to 1-x$. This asymmetry results from the matching
of matrix elements of the bare electromagnetic current between
the dressed LFWF in light-front QCD and of the
dressed electromagnetic current between hadronic wave
functions in AdS/QCD.

Concerning the $\bfk$ dependence of the 
$\varphi_q^{(1,2)}$ functions we use a specific 
functional form for them --- Gaussian ansatz. 
However, a generalized ansatz for $\varphi_q^{(i)}$ reads 
\eq 
\varphi_q^{(1)}(x,\bfk) &=& \frac{4 \pi}{M_N} \, 
\sqrt{\frac{q_v(x) + \delta q_v(x)}{2}} \, \sqrt{D_q^{(1)}(x)} \nonumber\\ 
&\times& \psi_1\biggl(- \frac{\bfk^2}{2 M_N^2} D_q^{(1)}(x) \biggr) \,, 
\nonumber\\
\varphi_q^{(2)}(x,\bfk) &=& \eta_q \, \frac{4 \pi}{M_N} \, 
\sqrt{\frac{q_v(x) - \delta q_v(x)}{2}} \, D_q^{(2)}(x) \nonumber\\ 
&\times& \psi_2\biggl(- \frac{\bfk^2}{2 M_N^2} D_q^{(2)}(x) \biggr) 
\en 
where the functions $\psi_1$ and $\psi_2$ must satisfy the normalization 
conditions following from Eq.~(\ref{norm_cond1}) 
\eq 
\int d^2\bfk \, 
\Big[\psi_1(-\bfk^2)\Big]^2 &=& \frac{\pi}{2} \,, \nonumber\\
\int d^2\bfk \, \bfk^2 \, 
\Big[\psi_2(-\bfk^2)\Big]^2 &=& \frac{\pi}{4} \,. 
\en  

\section{Light-front decompositions for the nucleon quantities} 
\label{Sec_LF_Decompositions}

\subsection{Form factors and parton distributions}

In this section we collect the well-known decompositions of the nucleon 
form factors and parton distributions in terms of the nucleon LFWFs. 
First we quote~\cite{Radyushkin:1998rt} the connection 
of the nucleon Dirac and Pauli form factors $F_{1,2}^N$ ($N=p,n$) with
the valence quark distributions $F_{1,2}^q$ ($q=u,d$) in nucleons with 
\eq\label{FF}
F_i^{p(n)}(Q^2) &=& \frac{2}{3} F_i^{u(d)}(Q^2) 
                 -  \frac{1}{3} F_i^{d(u)}(Q^2)\, . 
\en 
The valence quark distributions are related to the the nucleon nonforward 
parton densities (NPDs)~\cite{Radyushkin:1998rt} ${\cal H}^q(x,Q^2)$ and 
${\cal E}^q(x,Q^2)$ evaluated at zero skewness $\xi = 0$ as 
\eq\label{FF_NPD}
F_1^q(Q^2) &=& \int_{0}^{1} dx \, \mathcal{H}^{q}(x,Q^2)  \,,\nonumber\\
F_2^q(Q^2) &=& \int_{0}^{1} dx \, \mathcal{E}^{q}(x,Q^2)  \, ,
\en 
where $Q^2 = -q^2 > 0$ is the Euclidean momentum squared. 
At $Q^2=0$ the NPDs are related to the quark densities --- 
valence $q_v(x)$ and magnetic $\mathcal{E}_{q}(x)$ as 
\eq\label{norm_GPDs}
\mathcal{H}^{q}(x,0)=q_{v}(x)\,, 
\quad 
\mathcal{E}^{q}(x,0)=\mathcal{E}^{q}(x) \,, 
\en
which are normalized as 
\eq\label{normalization} 
n_q &=& F_1^q(0) = \int\limits_0^1 dx \, q_v(x)\,, \nonumber\\
\kappa_q &=& F_2^q(0) = \int\limits_0^{1} dx \, \mathcal{E}^q(x)\,, 
\en
where $\kappa_q$ is the anomalous quark magnetic moment. 

The nucleon Sachs form factors 
$G_{E/M}^N(Q^2)$ and the electromagnetic radii $\la r^2_{E/M} \ra^N$
are given in terms of the Dirac and Pauli form factors as   
\eq 
G_E^N(Q^2) &=& F_1^N(Q^2) - \frac{Q^2}{4m_N^2} F_2^N(Q^2)\,, 
\nonumber\\
G_M^N(Q^2) &=& F_1^N(Q^2) + F_2^N(Q^2)\,, \nonumber\\[3mm]
\la r^2_E \ra^N &=& - 6 \, \frac{dG_E^N(Q^2)}{dQ^2}\bigg|_{Q^2 = 0} \,, 
\nonumber\\
\la r^2_M \ra^N &=& - \frac{6}{G_M^N(0)} \, 
\frac{dG_M^N(Q^2)}{dQ^2}\bigg|_{Q^2 = 0}  \,, 
\en 
where $G_M^N(0) \equiv \mu_N$ is the nucleon magnetic moment. 

The light-front representation~\cite{LFQCD1,LFQCD2,LFQCD3,Brodsky_Drell} 
for the Dirac and Pauli quark form factors is  
\eq 
F_1^q(Q^2) &=& 
\int\limits_0^1 dx 
\int\frac{d^2\bfk}{16\pi^3} \, 
\biggl[ 
\psi_{+q}^{+\, \ast}(x,\bfk')\psi_{+q}^+(x,\bfk) 
\label{FF_LFQCD1} \nonumber\\
&+&\psi_{-q}^{+\, \ast}(x,\bfk')\psi_{-q}^+(x,\bfk) 
\biggr] \,, \\
F_2^q(Q^2) &=& - \frac{2M_N}{q^1-iq^2}
\int\limits_0^1 dx 
\int\frac{d^2\bfk}{16\pi^3} \nonumber\\
&\times&
\biggl[ 
\psi_{+q}^{+\, \ast}(x,\bfk')\psi_{+q}^-(x,\bfk) 
\nonumber\\
&+&\psi_{-q}^{+\, \ast}(x,\bfk')\psi_{-q}^-(x,\bfk) 
\biggr] \,, \label{FF_LFQCD2} 
\en  
where $\bfk' = \bfk + \bfq (1-x)$. 
Here $\psi_{\lambda_q q}^{\lambda_N}(x,\bfk)$ 
are the LFWFs at the initial scale $\mu_0$ with specific helicities for  
the nucleon $\lambda_N  = \pm$ and for the struck quark $\lambda_q = \pm $, 
where plus and minus correspond to $+\frac{1}{2}$ and $-\frac{1}{2}$, 
respectively. We work in the frame with $q=(0,0,\bfq)$, and where  
the Euclidean momentum squared is $Q^2 = \bfq^2$. For the initial scale 
we choose the value $\mu_0 \sim 1$ GeV which is used in the most 
of the global fits. 

The expressions for the quark helicity-independent NPDs $\mathcal{H}^{q}$ 
and $\mathcal{E}^{q}$ in the nucleon read 
\eq 
\mathcal{H}^q(x,Q^2) &=& 
\frac{q_v(x)+\delta q_v(x)}{2} \, e^{- t_q^{(11)}(x,Q^2)} \nonumber\\
&+& 
\frac{q_v(x)-\delta q_v(x)}{2} \, e^{- t_q^{(22)}(x,Q^2)} \nonumber\\
&\times& \Big[1 - t_q^{(22)}(x,Q^2)\Big]\,, \\ 
\mathcal{E}^q(x,Q^2) &=& {\cal E}^q(x) \, 
e^{- t_q^{(12)}(x,Q^2)} \,, 
\en 
where 
\eq 
\hspace*{-.5cm}
t_q^{(ij)}(x,Q^2) &=& \frac{Q^2}{4M_N^2} \, 
\frac{2 D_q^{(i)}(x) \, D_q^{(j)}(x)}
{D_q^{(i)}(x) + D_q^{(j)}(x)} \, (1-x)^2 \,. 
\en  
The magnetization PDF ${\cal E}^q(x)$ reads 
\eq 
\hspace*{-.4cm}
{\cal E}^q(x) = 4 \eta_q \, \sqrt{q_v^2(x) - \delta q_v^2(x)} \, 
\sqrt{D_q^{(1)}(x)} \, \frac{(1-x) \, \sigma_q(x)} 
{\Big[1 +\sigma_q(x)\Big]^2} \,. 
\en  
where $\sigma_q(x) = D_q^{(2)}(x)/D_q^{(1)}(x)$. 

Our expressions for the helicity-independent NPDs and PDFs 
contain only the four unknown functions $D_{q}^{(i)}(x)$ with 
$q = u, d$ and $i = 1, 2$. 

Note that by an appropriate choice of the longitudinal functions 
$D_{q}^{(i)}(x)$ we can guarantee the required scaling of the nucleon form 
factors at large $Q^2$. For example, if we adopt the following 
scaling of the quark helicity-independent PDFs:  
\eq\label{pdf_scalings} 
& &q_v(x) \sim (1-x)^{3}\,, \ \ 
\mathcal{E}^q(x)  \sim (1-x)^{5}\,, \ \ 
\en 
we should choose $D_q^{(i)}$ functions with the
scaling behavior 
\eq 
D_q^{(1)}(x) \sim (1-x)^0\,, \quad 
D_q^{(2)}(x) \sim (1-x)^{-1}\,. 
\en 
Thus we obtain the correct large $Q^2$ scaling of the quark form factors
of the form 
\eq\label{scaling_u1}
\hspace*{-.5cm}
F_1^q(Q^2) \sim \int\limits_0^1 dx (1-x)^3  
\exp\biggl[-\frac{Q^2}{4M_N^2}  (1-x)^2 
\biggr] \sim \frac{1}{Q^4}
\en
and 
\eq\label{scaling_u2}
\hspace*{-.5cm}
F_2^q(Q^2) \sim  \int\limits_0^1 dx (1-x)^5  
\exp\biggl[-\frac{Q^2}{4M_N^2}  (1-x)^2  
\biggr] \sim \frac{1}{Q^6} \,. 
\en 
This behavior guarantees the correct power scaling of the nucleon Dirac and 
Pauli form factors at higher $Q^2$ consistent with 
quark counting rules~\cite{Brodsky:1973kr,Matveev:1973kr}: 
\eq 
F_1^N(Q^2) \sim 1/Q^4\,, \quad 
F_2^N(Q^2) \sim 1/Q^6\,.  
\en 

\subsection{Transverse momentum-dependent parton distributions}

In the quark-diquark model, the light-front decomposition for the
transverse momentum-dependent parton distributions (TMDs) is discussed 
in detail in Ref.~\cite{Bacchetta:2008af} (see also Ref.~\cite{Maji:2015vsa}).
For recent progress in the extraction of TMDs from data,  
see e.g. Refs.~\cite{Bacchetta:2012ty}-\cite{Kang:2015msa}.   
The set of the valence quark $T$-even TMDs for the case 
of the quark-scalar diquark model is given by~\cite{Bacchetta:2008af}:  
\eq
\hspace*{-.75cm} 
& &f_1^{q_v}(x,\bfk) \equiv h_{1T}^{q_v}(x,\bfk) \nonumber\\
\hspace*{-.75cm} 
&=& 
\frac{1}{16 \pi^3} \, \biggl[ 
|\psi_{+q}^+(x,\bfk)|^2 
+ |\psi_{-q}^+(x,\bfk)|^2  \biggr] \nonumber\\
\hspace*{-.75cm} 
&=& 
\frac{1}{16 \pi^3} \, \biggl[ 
\Big(\varphi_q^{(1)}(x,\bfk)\Big)^2 
+ \frac{\bfk^2}{M_N^2}\Big(\varphi_q^{(2)}(x,\bfk)\Big)^2 \biggr]
\,,
\en
\eq
\hspace*{-.75cm} 
& &g_{1L}^{q_v}(x,\bfk) =
\frac{1}{16 \pi^3} \, \biggl[ 
|\psi_{+q}^+(x,\bfk)|^2 
- |\psi_{-q}^+(x,\bfk)|^2  \biggr]\nonumber\\
\hspace*{-.75cm} 
&=& 
\frac{1}{16 \pi^3} \, \biggl[ 
\Big(\varphi_q^{(1)}(x,\bfk)\Big)^2 
- \frac{\bfk^2}{M_N^2}\Big(\varphi_q^{(2)}(x,\bfk)\Big)^2 \biggr]
\,,
\en 
\eq 
\hspace*{-.5cm} 
& &g_{1T}^{q_v}(x,\bfk) \equiv - h_{1L}^{\perp q_v}(x,\bfk) \nonumber\\
\hspace*{-.75cm} 
&=& 
\frac{1}{16 \pi^3} \, \biggl[ 
\psi_{+q}^{+\, \ast}(x,\bfk) \psi_{+q}^-(x,\bfk) \, 
\frac{M_N}{k^1 - ik^2} \nonumber\\
\hspace*{-.75cm} 
&+& 
\psi_{+q}^{-\, \ast}(x,\bfk) \psi_{+q}^+(x,\bfk) \, 
\frac{M_N}{k^1 + ik^2} 
\biggr]\nonumber\\
\hspace*{-.75cm} 
&=& 
\frac{1}{8 \pi^3} \, \varphi_q^{(1)}(x,\bfk) 
                  \, \varphi_q^{(2)}(x,\bfk) \,, 
\en 
\eq
\hspace*{-.75cm} 
& &h_1^{q_v}(x,\bfk) \equiv h_{1T}^{q_v}(x,\bfk)  
+ \frac{\bfk^2}{2M_N^2} h_{1T}^{\perp q_v}(x,\bfk) \nonumber\\
\hspace*{-.75cm} 
&=& \frac{1}{2} \Big[f_1^{q_v}(x,\bfk) + g_{1L}^{q_v}(x,\bfk) \Big] \nonumber\\
\hspace*{-.75cm} 
&=& \frac{1}{16 \pi^3} \, |\psi_{+q}^+(x,\bfk)|^2 =\frac{1}{16 \pi^3} \, 
\Big(\varphi_q^{(1)}(x,\bfk)\Big)^2 
\,,
\en 
\eq 
\hspace*{-.75cm}
& &\frac{\bfk^2}{2M_N^2} h_{1T}^{\perp q_v}(x,\bfk) =
\frac{1}{2} \Big[g_{1L}^{q_v}(x,\bfk) - f_1^{q_v}(x,\bfk)\Big] \nonumber\\
\hspace*{-.75cm} 
&=&  g_{1L}^{q_v}(x,\bfk) - h_1^{q_v}(x,\bfk)\nonumber\\
\hspace*{-.75cm}
&=& - \frac{1}{16 \pi^3} \, |\psi_{-q}^+(x,\bfk)|^2 \nonumber\\ 
&=& - \frac{1}{16 \pi^3} \, \frac{\bfk^2}{M_N^2} \,  
\Big(\varphi_q^{(2)}(x,\bfk)\Big)^2 \,. 
\en 
Using our expressions for the LFWFs we can express the TMDs 
through the PDFs 
\eq 
f_1^{q_v}(x,\bfk) &\equiv& h_{1T}^{q_v}(x,\bfk) \nonumber\\
&=&{\cal F}_1(x,\bfk) + {\cal F}_2(x,\bfk)\,, \nonumber\\ 
g_{1L}^{q_v}(x,\bfk) &=& {\cal F}_1(x,\bfk) - {\cal F}_2(x,\bfk)\,, 
\nonumber\\ 
g_{1T}^{q_v}(x,\bfk) &\equiv& - h_{1L}^{\perp q_v}(x,\bfk) = 
{\cal F}_3(x,\bfk) \,, \nonumber\\
h_1^{q_v}(x,\bfk) &=& {\cal F}_1(x,\bfk)\,, 
\nonumber\\
\frac{\bfk^2}{2M_N^2} h_{1T}^{\perp q_v}(x,\bfk) &=& 
- {\cal F}_2(x,\bfk)\,, 
\en 
where 
\eq 
\vspace*{-.2cm}
{\cal F}_1(x,\bfk) &=& \frac{1}{\pi M_N^2} \, 
\frac{q_v(x) + \delta q_v(x)}{2} \, D_q^{(1)}(x) \nonumber\\ 
\vspace*{-.2cm}&\times&e^{- \frac{\bfk^2}{M_N^2} D_q^{(1)}(x)}\,,  \nonumber\\ 
{\cal F}_2(x,\bfk) &=& \frac{1}{\pi M_N^2} \, 
\frac{q_v(x) - \delta q_v(x)}{2} \, 
\frac{\bfk^2}{M_N^2}
\Big(D_q^{(2)}(x)\Big)^2 \nonumber\\
\vspace*{-.2cm}&\times&e^{- \frac{\bfk^2}{M_N^2} D_q^{(2)}(x)}  
\,, \nonumber\\
\vspace*{-.2cm}{\cal F}_3(x,\bfk) &=& \eta_q \,  
\sqrt{\frac{4M_N^2}{\bfk^2} \, 
{\cal F}_1(x,\bfk) \, {\cal F}_2(x,\bfk) } \nonumber\\
\vspace*{-.2cm}&=&\frac{1}{\pi M_N^2} \, \eta_q \, 
\sqrt{q_v^2(x) - \delta q_v^2(x)} \, \sqrt{D_q^{(1)}(x)} \, 
D_q^{(2)}(x) \nonumber\\ 
\vspace*{-.2cm}&\times&e^{- \frac{\bfk^2}{2 M_N^2} \, 
\Big(D_q^{(1)}(x) + D_q^{(2)}(x)\Big) }\,.  
\en 
Performing the $\bfk$-integration over the TMDs with 
\eq 
{\rm TMD}(x) &=& \int d^2\bfk \, {\rm TMD}(x,\bfk) \,, \nonumber\\
\overline{{\rm TMD}}(x) &=& \int d^2\bfk \, \frac{\bfk^2}{2M_N^2} \, 
{\rm TMD}(x,\bfk) 
\en 
results in the identities
\eq 
f_1^{q_v}(x) &\equiv& h_{1T}^{q_v}(x) \, = \, q_v(x)\,, 
\nonumber\\[2mm]
g_{1L}^{q_v}(x) &=& \delta q_v(x)\,, 
\nonumber\\[2mm]
g_{1T}^{q_v}(x) &\equiv& - h_{1L}^{\perp q_v}(x) = 
{\cal E}^q(x) \ \frac{1 + \sigma_q(x)}{2 (1-x)} \,, 
\nonumber\\[2mm]
h_1^{q_v}(x) &=& \frac{q_v(x) + \delta q_v(x)}{2}\,, \nonumber\\ 
\overline{h_{1T}^{\perp q_v}}(x) &=& 
- \frac{q_v(x) - \delta q_v(x)}{2}\,. 
\en 
Finally, the integration over $x$ leads to the normalization conditions
\eq 
\int\limits_0^1 dx f_1^{q_v}(x) = \int\limits_0^1 dx h_{1T}^{q_v}(x) = n_q \,, 
\nonumber\\ 
\int\limits_0^1 dx g_{1L}^{q_v}(x) = g_A^q \,, 
\nonumber\\ 
\int\limits_0^1 dx h_1^{q_v}(x) = g_T^q \,, 
\en 
where $g_T^q$ is the tensor charge.

Our TMD, independently on the longitudinal
functions $D_q^{(i)}$, satisfy all relations and inequalities found before
in theoretical approaches (see detailed discussion in 
Refs.~\cite{Bacchetta:1999kz,Avakian:2010br,Lorce:2011zta,%
Soffer:1994ww,Maji:2015vsa}. 
In particular, our TMDs in agreement with QCD and 
other models~\cite{Bacchetta:1999kz,Avakian:2010br} 
(see also Ref.~\cite{Maji:2015vsa}) satisfy the following 
inequality relations:  
\eq\label{inequalities} 
\hspace*{-.6cm}
& &f_1^{q_v}(x,\bfk) > 0 \,, \nonumber\\
\hspace*{-.6cm}
& &|g_{1L}^{q_v}(x,\bfk)| \le |f_1^{q_v}(x,\bfk)| \,, \nonumber\\
\hspace*{-.6cm}
& &|h_1^{q_v}(x,\bfk)| \le |f_1^{q_v}(x,\bfk)| \,, \nonumber\\
\hspace*{-.6cm}
& &|g_{1T}^{q_v}(x,\bfk)| \le |F_1^{q_v}(x,\bfk)| = 
\sqrt{\frac{M_N^2}{\bfk^2}} \, 
|f_1^{q_v}(x,\bfk)| 
\en 
which follow from the simple positivity condition for 
our functions ${\cal F}_1(x,\bfk)$ and ${\cal F}_2(x,\bfk)$ 
\eq 
\Big[{\cal F}_1(x,\bfk) - {\cal F}_2(x,\bfk)\Big]^2 \geq 0 \,. 
\en 
Additionally, we confirm the inequality between the tensor and axial 
charges found in lattice QCD and different model (see discussion in 
Refs.~\cite{Efremov:2009ze} and~\cite{Maji:2015vsa}) and the generalized 
inequality 
\eq 
|h_1^{q_v}(x,\bfk)| \ge |g_{1L}^{q_v}(x,\bfk)|
\en 
observed before in the framework of parton model~\cite{Efremov:2009ze} 
and derived recently in the quark-diquark model in Ref.~\cite{Maji:2015vsa}. 
Finally, our TMDs satisfy the non-linear relation found in 
Ref.~\cite{Efremov:2009ze} and recently confirmed in Ref.~\cite{Maji:2015vsa}: 
\eq 
h_1^{q_v}(x,\bfk) h_{1T}^{\perp q_v}(x) = - \frac{1}{2} \, 
\Big[h_{1L}^{\perp q_v}(x)\Big]^2 \,.
\en 
We would like to stress that the last inequality condition 
in Eq.~(\ref{inequalities}) relating $g_{1T}^{q_v}$ and $f_1^{q_v}$ 
after integration over $\bfk$ is also 
fulfilled in our approach. In particular, after integration 
over $\bfk$ we get 
\eq\label{Ineq_g1T_f1} 
\hspace*{-.7cm}
g_{1T}^{q_v}(x) &=& {\cal E}^q(x) \ \frac{1 + \sigma_q(x)}{2 (1-x)}  
\le F_1^{q_v}(x) = \sqrt{\pi \, D_q^{(1)}(x)}\nonumber\\
\hspace*{-.7cm}
&\times&\biggl[
\frac{q_v(x) + \delta q_v(x)}{2} 
+ \sigma_q(x) \, \frac{q_v(x) - \delta q_v(x)}{4}\biggr] \,. 
\en 
The inequality~(\ref{Ineq_g1T_f1}) is fulfilled because it is reduced to 
more trivial inequality 
\eq 
[1 + \sigma_q(x)]^2 > \frac{8}{\pi} \, \sqrt{\sigma_q(x)} \,, 
\en 
which occurs because of 
\eq 
[1 + \sigma_q(x)]^2 > \sqrt{8 \, \sigma_q(x)} 
\en 
and 
\eq 
\sqrt{8} > \frac{8}{\pi} \,. 
\en 
In Sect.~\ref{Sec_Results} we present a plot where we compare 
our predictions for the $g_{1T}^{q_v}(x)$ TMDs with corresponding 
upper limits defined by right-hand side of Eq.~(\ref{Ineq_g1T_f1}).  

\subsection{Wigner distributions} 

In light-front QCD the 
Wigner distributions 
read~\cite{Ji:2003ak,Belitsky:2003nz,Meissner:2008ay,%
Meissner:2009ww,Chakrabarti:2016yuw}  
\eq 
\rho^{q[\Gamma]}(x,\bfb,\bfk;S) &=& \int \frac{d^2\bfD}{4\pi^2} \, 
e^{- i \bfD \bfb} \nonumber\\
&\times& W^{q[\Gamma]}(x,\bfD,\bfk;S)\,, 
\en 
where $W^{q[\Gamma]}(x,\bfD,\bfk;S)$ is the matrix element of 
the Wigner operator for $\Delta^+ = 0$ and $z^+ = 0$. 
The light-front decomposition of the Wigner matrix elements 
$W^{q[\Gamma]}(x,\bfD,\bfk;S)$ is given by~\cite{Meissner:2008ay}  
\eq 
\hspace*{-.75cm}
& &W^{q[\gamma^+]}(x,\bfD,\bfk;\pm e_z)\nonumber\\
\hspace*{-.75cm}
&=&
\frac{1}{16 \pi^3} \, \biggl[ 
\psi^{\pm \dagger}_{q+}(x,\bfk^+) \psi^{\pm}_{q+}(x,\bfk^-) 
+ 
\psi^{\pm \dagger}_{q-}(x,\bfk^+) \psi^{\pm}_{q-}(x,\bfk^-) 
\biggr]
\,, \nonumber\\
\hspace*{-.75cm}
& &\\
\hspace*{-.75cm}
& &W^{q[\gamma^+\gamma^5]}(x,\bfD,\bfk;\pm e_z)\nonumber\\
\hspace*{-.75cm}
&=& 
\frac{1}{16 \pi^3} \, \biggl[ 
\psi^{\pm \dagger}_{+q}(x,\bfk^+) \psi^{\pm}_{+q}(x,\bfk^-) 
- 
\psi^{\pm \dagger}_{-q}(x,\bfk^+) \psi^{\pm}_{-q}(x,\bfk^-) 
\biggr]
\,, \nonumber
\en 
where $\bfk^\pm = \bfk \pm (1-x) \bfD/2$. 

Next we use the standard definitions of the Wigner distributions,  
specified by the nucleon helicity $\lambda_N$ and the quark 
helicity $\lambda_q$~\cite{Meissner:2008ay}   
\eq 
\hspace*{-.5cm}
\rho^q_{\lambda_N\lambda_q}(x,\bfb,\bfk) &=& \frac{1}{2} \biggl[ 
\rho^{q[\gamma^+]}(x,\bfb,\bfk,\lambda_N \vec{e}_z) \nonumber\\
\hspace*{-.5cm}
&+& \lambda_q \rho^{q[\gamma^+\gamma^5]}(x,\bfb,\bfk,\lambda_N \vec{e}_z) 
\biggr]\,,
\en 
which can be further expressed in terms of distributions 
where the proton and the struck quark are unpolarized ($U$) or 
longitudinally polarized ($L$): 
\eq 
\rho^q_{\lambda_N\lambda_q} &=& \frac{1}{2} 
\biggl[ \rho^q_{UU} + \lambda_N \rho^q_{LU} + 
\lambda_q \rho^q_{UL} + \lambda_N \lambda_q \rho^q_{LL} 
\biggr]\,, \nonumber\\
\rho^q_{UU} &=& \rho^{q[\gamma^+]}(\vec{e}_z) + 
\rho^{q[\gamma^+]}(-\vec{e}_z)\,, 
\nonumber\\
\rho^q_{LU} &=& \rho^{q[\gamma^+]}(\vec{e}_z) 
- \rho^{q[\gamma^+]}(-\vec{e}_z)\,, 
\nonumber\\
\rho^q_{UL} &=& \rho^{q[\gamma^+\gamma^5]}(\vec{e}_z) 
+ \rho^{q[\gamma^+\gamma^5]}(-\vec{e}_z)\,, 
\nonumber\\
\rho^q_{LL} &=& \rho^{q[\gamma^+\gamma^5]}(\vec{e}_z) 
- \rho^{q[\gamma^+\gamma^5]}(-\vec{e}_z)\, .
\en 
The Fourier transforms $\omega_{AB}$, $A,B=U,L$ 
with respect to the $\bfb$ variable we define as 
\eq 
\hspace*{-.2cm}\omega_{AB}^q(x,\bfD,\bfk) = \int d^2\bfb \, 
e^{i \bfD \bfb} \, \rho_{AB}^q(x,\bfb,\bfk)\,.  
\en 
The expressions for the Wigner distributions $\rho^q_{AB}$ 
and $\omega^q_{AB}$ in the light-front quark-diquark approach are listed 
in Appendix~\ref{AppendixA}.   

\subsection{Quark orbital angular momentum} 

Following Ji~\cite{Ji:1996ek} we define the quark contribution 
to the nucleon angular momentum:  
\eq 
J_z^q &=& L_z^q + S_z^q \nonumber\\
&=& \frac{1}{2} \int\limits_0^1 dx x 
\, \Big[ {\cal H}^q(x,0) + {\cal E}^q(x,0) \Big] \nonumber\\
&=& \frac{1}{2} \int\limits_0^1 dx x \Big[ q_v(x) + {\cal E}^q(x) \Big]\,, 
\en 
where the quark orbital angular momentum (OAM) $L_z^q$ 
and internal spin $S_z^q$ contributions are defined as 
\eq 
L_z^q &=& \frac{1}{2} \int\limits_0^1 dx \, \biggl( x 
\, \Big[ {\cal H}^q(x,0) + {\cal E}^q(x,0) \Big] - 
\tilde H^q(x,0) \biggr) \nonumber\\
&=& \frac{1}{2} \int\limits_0^1 dx \biggl( 
x \Big[ q_v(x) + {\cal E}^q(x) \Big] - \delta q_v(x) \biggr) 
\en 
and 
\eq 
S_z^q = \frac{1}{2} \int\limits_0^1 dx \, \tilde H^q(x,0) 
= \frac{1}{2} \int\limits_0^1 \, \delta q_v(x) \,.   
\en 
Integrating the TMD $- \frac{\bfk^2}{2M_N^2} \, h_{1T}^{\perp q_v}(x,\bfk)$ 
over $x$ and $\bfk$ one can derive the quantity ${\cal L}_z^q$ 
\eq
{\cal L}_z^q &=& - \int\limits_0^1 dx \int d^2\bfk \, \frac{\bfk^2}{2M_N^2}
\, h_{1T}^{\perp q_v}(x,\bfk) 
\en 
which is some quark models~\cite{Avakian:2008dz,%
Avakian:2009jt,Avakian:2010br,She:2009jq} is equal 
to the quark OAM, but in general, in a gauge theory, it is not the case 
and ${\cal L}_z^q \neq L_z^q$ (see discussion in Refs.~\cite{Burkardt:2008ua} 
and~\cite{Lorce:2011kd}).  
In particular, in our approach the quantity ${\cal L}_z^q$  
is not related to the quark OAM $L_z^q$  
\eq
{\cal L}_z^q &=& 
\frac{1}{2}
\int\limits_0^1 dx \,\Big( q_v(x) - \delta q_v(x) \Big) = 
\frac{n_q - g_A^q}{2} \nonumber\\
&\neq& L_z^q \,. 
\en 
Using $n_u = 2$ and $n_d = 1$ we get relation between 
quantities ${\cal L}_z^q$ and $S_z^q$: 
\eq\label{Lz_Sz} 
{\cal L}_z^u = 1 -  S_z^u\,, \ 
{\cal L}_z^d = \frac{1}{2} -  S_z^d\,. 
\en 
The next interesting quantity is the averaged quark orbital angular 
momentum (OAM) in a nucleon which is polarized in the 
$z$-direction~\cite{Meissner:2008ay,Lorce:2011kd,Lorce:2011ni}:  
\eq 
l_z^q &=& \la \hat{L}_z^q \ra^{[\gamma^+]}(\vec{e}_z) \nonumber\\
&=& 
\int\limits_0^1 dx \int d^2\bfk d^2\bfb \, (\bfb \times \bfk)_z 
\nonumber\\
&\times& \rho^{[\gamma^+]}(\bfb,\bfk,x,\vec{e}_z) \nonumber\\
&=&
\int\limits_0^1 dx \int d^2\bfk d^2\bfb \, (\bfb \times \bfk)_z 
\nonumber\\
&\times& \rho_{LU}^q(\bfb,\bfk,x) \nonumber\\
&=& \frac{1}{2} 
\int\limits_0^1 dx (1-x) \Big( q_v(x) - \delta q_v(x) \Big) \,. 
\en 
One can see that the $l_z^q$ is related with TMD $h_{1T}^{\perp q_v}(x,\bfk)$ 
by the integral representation over $x$ and $\bfk$ as 
\eq 
l_z^q &=& - \int\limits_0^1 dx \int d^2\bfk \, \frac{\bfk^2}{2M_N^2}
\, (1-x) \, h_{1T}^{\perp q_v}(x,\bfk) \,.
\en
Another relevant quantity is the correlation between the quark spin 
and the orbital angular 
momentum~\cite{Meissner:2008ay,Lorce:2011kd,Lorce:2011ni} with 
\eq 
C_z^q &=& 
\int\limits_0^1 dx \int d^2\bfk d^2\bfb \, (\bfb \times \bfk)_z 
\nonumber\\
&\times& \rho_{UL}^q(\bfb,\bfk,x) \,, 
\en 
which in quark-scalar diquark model~\cite{Kanazawa:2014nha} 
is opposite to the quantity $l_z^q$ with
\eq
C_z^q \equiv - l_z^q \,
\en 
because of 
$\rho_{UL}^q(\bfb,\bfk,x) = - \rho_{LU}^q(\bfb,\bfk,x)$.
It is also confirmed in our calculations. 

\subsection{Husimi distribution}

Finally, we consider the Husimi distribution function for the nucleon, 
which was recently discussed in detail 
in Refs.~\cite{Hagiwara:2014iya,Hatta:2015ggc}. 
As was stressed in~\cite{Hagiwara:2014iya,Hatta:2015ggc} 
this distribution is better behaved and positive in comparison to the 
Wigner distribution. It also gives a probabilistic interpretation 
and can be used to define the entropy of the nucleon 
as a measure of the complexity of the partonic structure. 
It also could be connected to the color glass condensate approach 
at small $x$. 

The Husimi distribution $h^q_{AB}(x,\bfb,\bfk)$ is defined as 
the integral of the Wigner distribution $\rho^q_{AB}(x,\bfb,\bfk)$ 
over the impact parameter~$\bfb$ and the transverse momentum~$\bfk$ 
\eq 
h^q_{AB}(x,\bfb,\bfk) &=& \frac{1}{\pi^2} 
\, \int d^2\bfk' d^2\bfb' \, e^{-(\bfb-\bfb')^2/l^2} 
\nonumber\\ 
&\times& e^{-l^2 (\bfk-\bfk')^2} \, \rho^q_{AB}(x,\bfb',\bfk') \,,
\en 
where $1/l^2 = \la \bfk^2 \ra$ is the average transverse momentum 
squared. 

Note that the double moment of the Husimi distribution 
$h^q_{UU}$ and $h^q_{LL}$ is the ordinary PDF: 
\eq 
f_1^{q_v}(x) &=& \int d^2\bfb d^2\bfk \, h^q_{UU}(x,\bfb,\bfk) \nonumber\\
&=& \int d^2\bfb  d^2\bfk \, \rho^q_{UU}(x,\bfb,\bfk) \,, \\
g_{1L}^{q_v}(x) &=& \int d^2\bfb d^2\bfk \, h^q_{LL}(x,\bfb,\bfk) \nonumber\\
&=& \int d^2\bfb  d^2\bfk \, \rho^q_{LL}(x,\bfb,\bfk) \,. 
\en 
In case of $h^q_{UL}$ and $h^q_{LU}$ the double moments equal zero. 

In quantum mechanics the Husimi distribution $h_{\rm QM}$ is positive 
definite and one can define the so-called the Husimi-Wahrl 
entropy~\cite{Wehrl:1978zz}, which in our case can be extended to 
define the entropy of the nucleon~\cite{Hagiwara:2014iya}
\eq 
S(x) &=& - \int d^2\bfb d^2\bfk h(x,\bfb,\bfk) \nonumber\\
&\times&\log\Big[h(x,\bfb,\bfk)\Big] \,. 
\en   
In particular, it is convenient to define two quantities 
\eq 
S^q_\pm(x) &=& - \int d^2\bfb d^2\bfk h^q_{\pm}(x,\bfb,\bfk) \nonumber\\ 
&\times&\log\Big[h^q_{\pm}(x,\bfb,\bfk)\Big] \,, 
\en 
where $h^q_{\pm} = (h^q_{UU} \pm h^q_{LL})/2$. 
Expressions for $S^q_\pm(x)$ are listed in the Appendix~\ref{AppendixA}.    

\section{Results} 
\label{Sec_Results}

In this paper we do not pretend on a precise analysis of the available 
nucleon data. Instead we first would like to illustrate how our method works. 
For this purpose we use the results for the NLO helicity-independent 
and helicity-dependent parton distributions 
at $\mu_{\rm NLO}^2~=~0.40$~GeV$^2$ 
from Refs.~\cite{Gluck:1998xa} and~\cite{Gluck:2000dy} as input: 
\eq
q(x) &=& q_v(x) + \bar q(x)\,, \, 
\delta q(x) \ = \  \delta q_v(x) + \delta \bar q(x)\,,
\nonumber\\
x u_v(x) &=& 0.632 \, x^{0.43} \, (1 - x)^{3.09} \, (1 + 18.2x) \,, \nonumber\\
d_v(x) &=& 0.394 \, (1 - x) \,  u_v(x) \,, \nonumber\\
x (\bar u + \bar d)(x) &=& 1.24 \, x^{0.20} \, (1 - x)^{8.5} \, 
(1 - 2.3 \sqrt{x} + 5.7 x) \,, \nonumber\\ 
x (\bar d - \bar u)(x) &=& 0.2 \, x^{0.43} \, (1 - x)^{12.4} \, 
(1 - 13.3 \sqrt{x} + 60 x) \,, \nonumber\\
x \delta u_v(x) &=& 2.043 \, x^{0.97} \, (1 - x)^{0.64} \, u(x)
\,, \nonumber\\
x \delta d_v(x) &=& - 2.709 \, x^{1.26} \, (1 - x)^{1.06} \, d(x)
\,, \nonumber\\
x \delta \bar u(x) &=& 1.727 \, x^{0.73} \, (1 - x)^{2.00} \, \bar u(x)
\,, \nonumber\\
\delta \bar d(x) &=& \delta \bar u(x) \frac{\delta u(x)}{\delta d(x)}
\,.
\en
The $D_q^{(i)}(x)$ are specified as 
\eq
D_q^{(1)}(x) &=& A_q \, \log({1/x}) \, (x+0.001)^{\alpha_q} \, 
(1-x)^{\beta_q}\,, 
\nonumber\\
\sigma_q(x)  &=& N_q \, e^{- \gamma_q x} \, 
x^{\bar\alpha_q} \, (1-x)^{\bar\beta_q}\,, 
\en 
where 
\eq 
& &A_u = 6.3385\,, \ A_d = 1.17396\,, \nonumber\\
& &\alpha_u = 0.37\,, \ \alpha_d = -0.31 \,, 
\, \beta_u = 0.09\,, \ \beta_d = -0.50 \,, \nonumber\\
& &N_u = 12.6\,, \ N_d = 2.8\,,  
\, \gamma_u = 3.70\,, \ \gamma_d = 0.45 \,, \nonumber\\
& &\bar\alpha_u = 0.045\,, \ \bar\alpha_d = 0 \,, 
\, \bar\beta_u = -0.60\,, \ \bar\beta_d = 0 \,. 
\en 
In Table I we present our results for the valence quark properties 
($J_z^q$, $L_z^q$, ${\cal L}_z^q$, $l_z^q$, $C_z^q$, $\kappa_q$)   
and compare them to results of other calculations 
(light-cone constituent quark model (LCCQM) and chiral quark-soliton 
model ($\chi$QSM))~\cite{Lorce:2011kd}. One can see that most of our 
results are different from the predictions of the 
LCCQM and $\chi$QSM approaches. This is caused by the difference in 
the magnetizazion PDFs ${\cal E}^q(x)$ (anomalous quark magnetic moments 
$\kappa_q$), helicity-dependent PDFs $\delta q_v(x)$ 
(quark contributions to internal spin $S_z^q$). Note that our 
magnetization PDFs are consistent with data for nucleon electromagnetic 
form factors and helicity-dependent PDFs $\delta q_v(x)$ are taken 
from Refs.~\cite{Gluck:1998xa,Gluck:2000dy}. 
Also we would like to stress that our results for 
the quantities ${\cal L}_z^q$ are clearly understood because 
they are related to the quantities $S_z^q$ by the relations~(\ref{Lz_Sz}). 
 
\begin{table}[htb]
\begin{center}
\caption{Valence quark properties.} 

\vspace*{.1cm}

\def\arraystretch{1.4}
\hspace*{-.4cm}
\begin{tabular}{|c|c|c|c|}
\hline
Quantity   &  LCCQM~\cite{Lorce:2011kd} & $\chi$QCM~\cite{Lorce:2011kd} & Our \\
\hline
$J_z^u$    & 0.569  & 0.566 & 0.358 \\
\hline
$J_z^d$    & $-$ 0.069  & $-$ 0.066     & $-$ 0.010 \\
\hline
$L_z^u$    & 0.071  & $-$ 0.008 & 0.055 \\
\hline
$L_z^d$    & 0.055  & 0.077     & $-$ 0.001 \\
\hline
$S_z^u$    & 0.498  & 0.574 & 0.303 \\
\hline
$S_z^d$    & $-$ 0.124  & $-$ 0.143 & $-$ 0.009 \\
\hline
${\cal L}_z^u$    & 0.169 & 0.093 & 0.697 \\
\hline
${\cal L}_z^d$    & $-$ 0.042 & $-$ 0.023 & 0.509 \\
\hline
$l_z^u$    & 0.131  & 0.073     & 0.598 \\
\hline
$l_z^d$    & -0.005 & $-$ 0.004 & 0.404 \\ 
\hline
$C_z^u$    & 0.227  & 0.130 & $-$ 0.598 \\
\hline
$C_z^d$    & 0.187  & 0.109 & $-$ 0.404 \\
\hline 
$\kappa_u$ & 1.867  & 1.766 & 1.673     \\ 
\hline 
$\kappa_d$ & $-$ 1.579  & $-$ 1.551 & $-$ 2.033     \\ 
\hline
\end{tabular}
\end{center}
\end{table}

\begin{figure}[ht!]
\begin{center}

\epsfig{figure=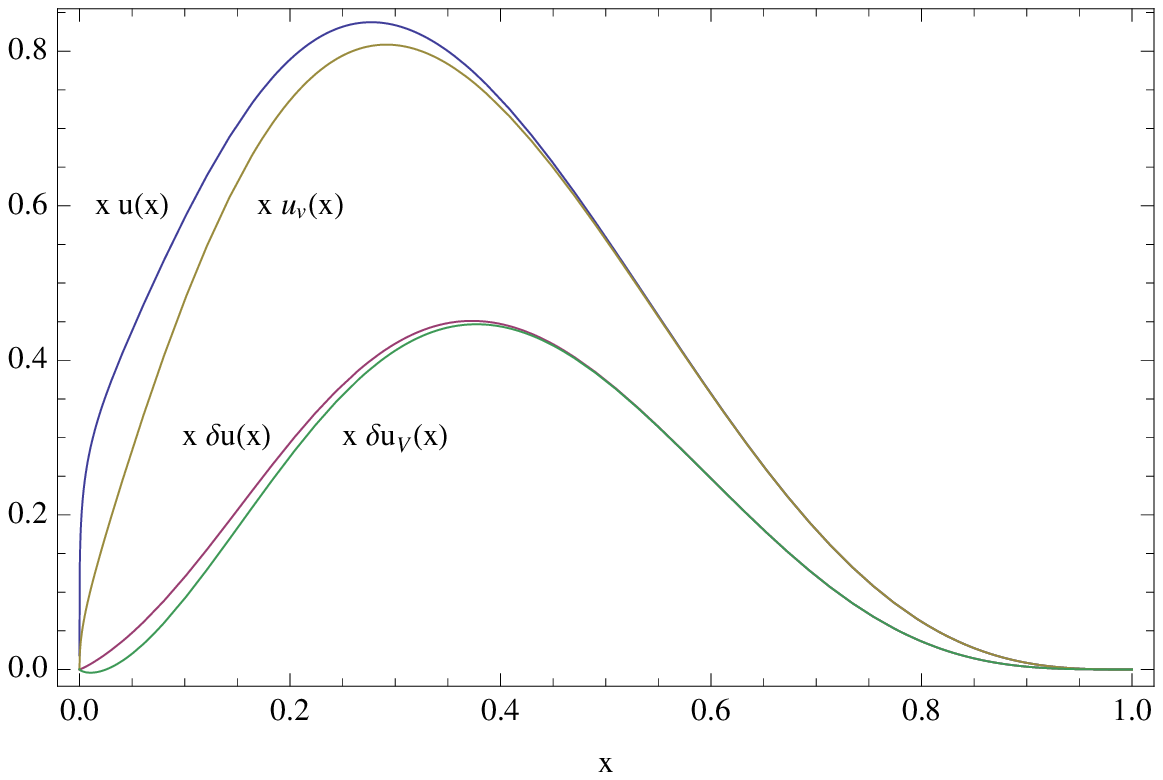,scale=.57}
\end{center}
\vspace*{-.6cm}
\noindent
\caption{$u$ quark PDFs multiplied with $x$.
\label{fig:uPDF}}

\begin{center}
\epsfig{figure=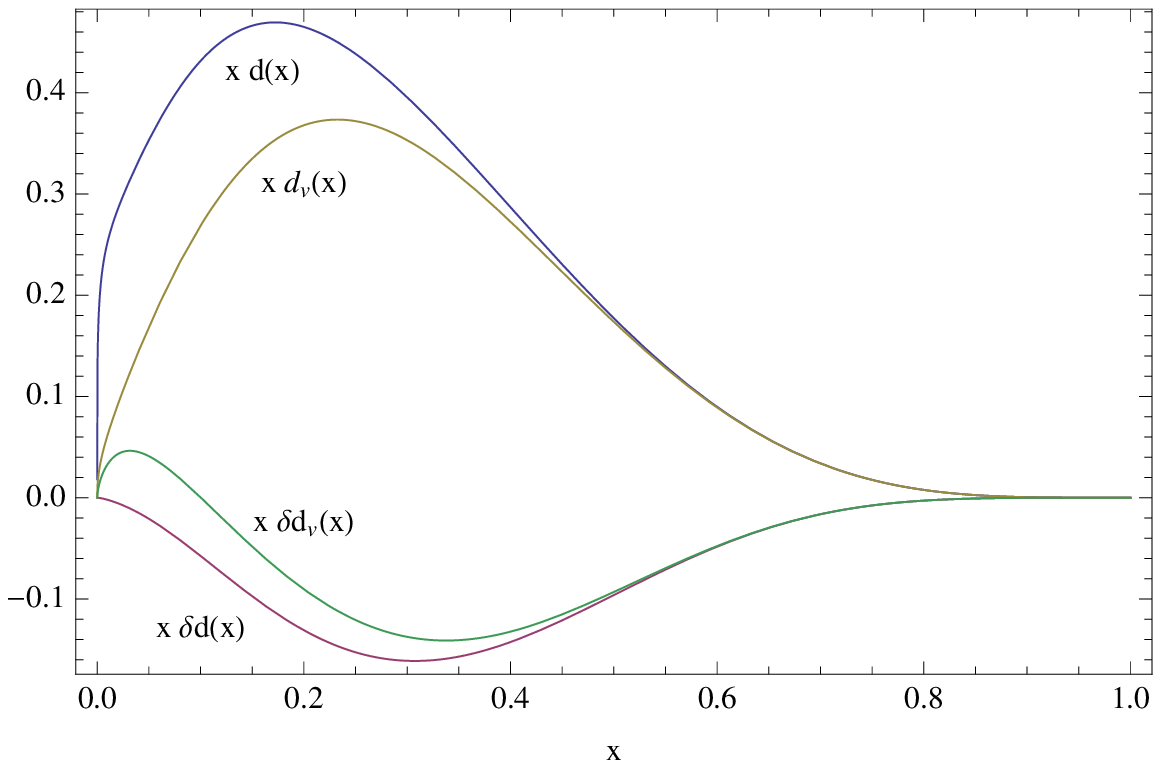,scale=.57}
\end{center}
\vspace*{-.6cm}
\noindent
\caption{$d$ quark PDFs multiplied with $x$.
\label{fig:dPDF}}

\begin{center}
\epsfig{figure=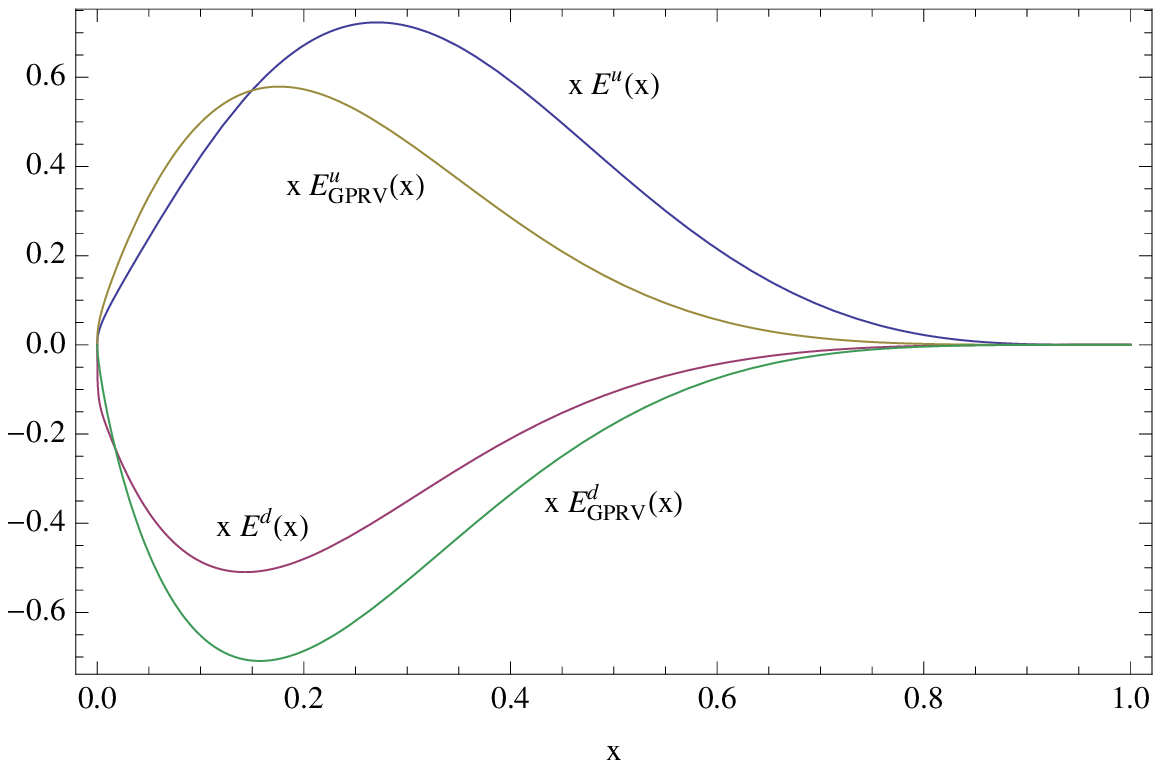,scale=.57}
\end{center}
\vspace*{-.6cm}
\noindent
\caption{quark magnetization PDFs: 
$x {\cal E}^q$ - our results, 
$x {\cal E}^q_{GPRV}$ - results of Ref.~\cite{Guidal:2004nd}. 
\label{fig:EqPDF}}

\begin{center}
\epsfig{figure=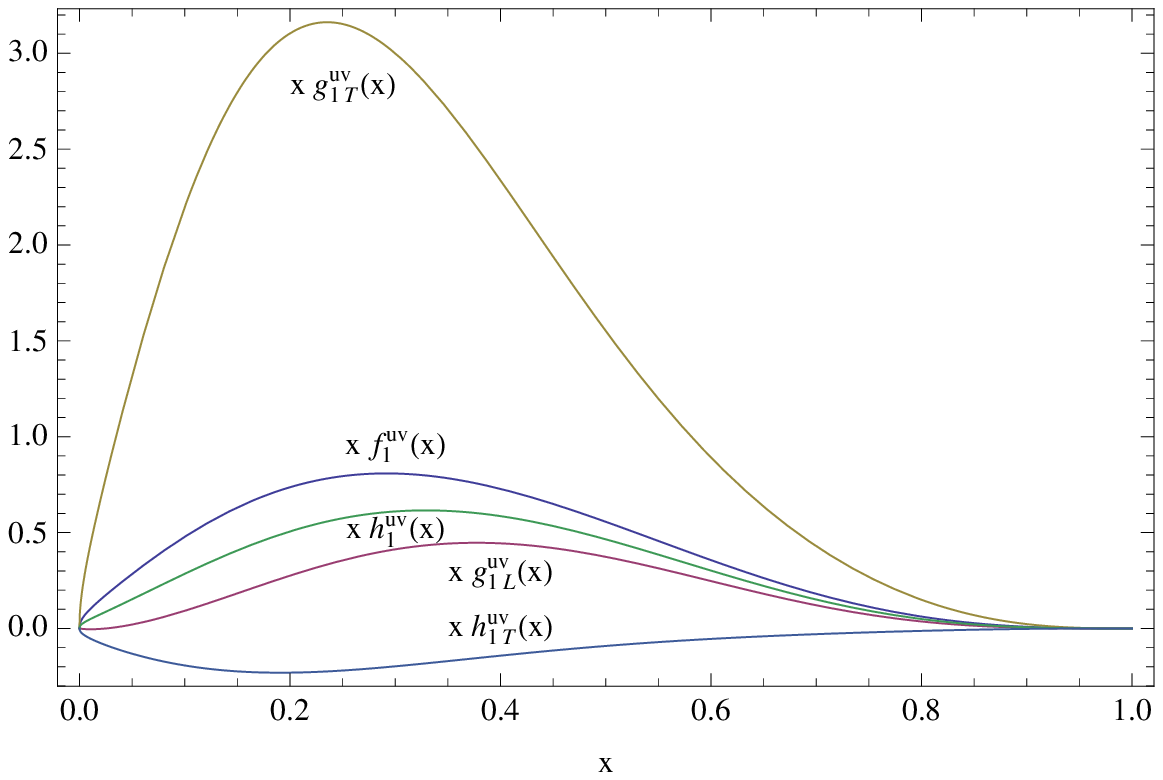,scale=.57}
\end{center}
\vspace*{-.6cm}
\noindent
\caption{$u$ quark TMDs multiplied with $x$.
\label{fig:hux}}

\end{figure}

\begin{figure}

\begin{center}
\epsfig{figure=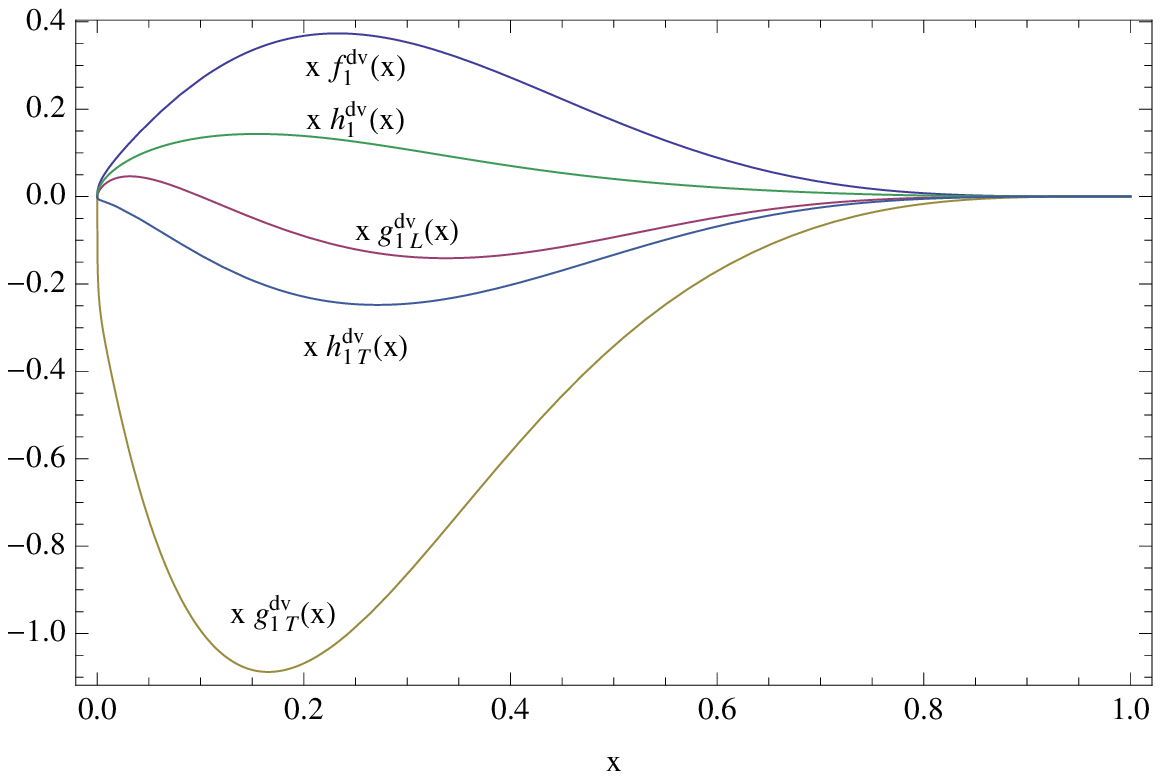,scale=.57}
\end{center}
\vspace*{-.5cm}
\noindent
\caption{$d$ quark TMDs multiplied with $x$.
\label{fig:hdx}}

\begin{center}
\epsfig{figure=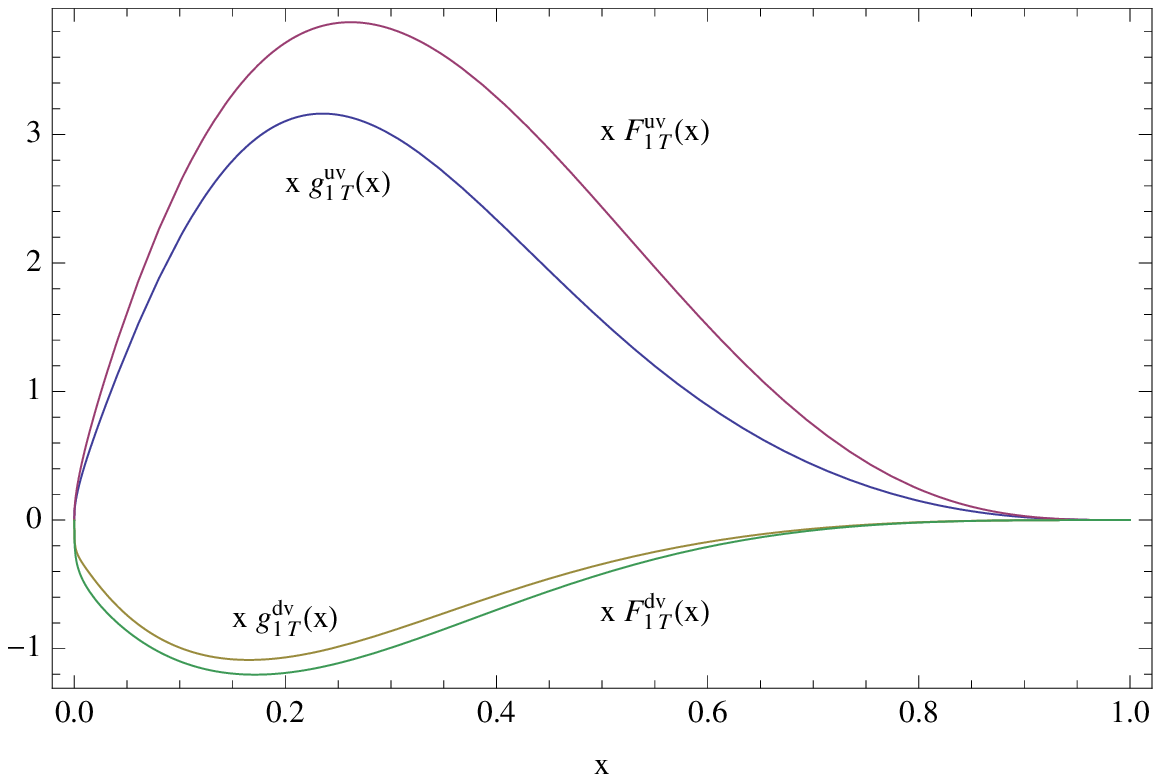,scale=.57}
\end{center}
\vspace*{-.5cm}
\noindent
\caption{Comparison our predictions for 
$x g_{1T}^{q_v}(x)$ quark TMDs multiplied with 
corresponding upper limits $x F_{1}^{q_v}(x)$.
\label{fig:g1TLimits}}

\begin{center}
\epsfig{figure=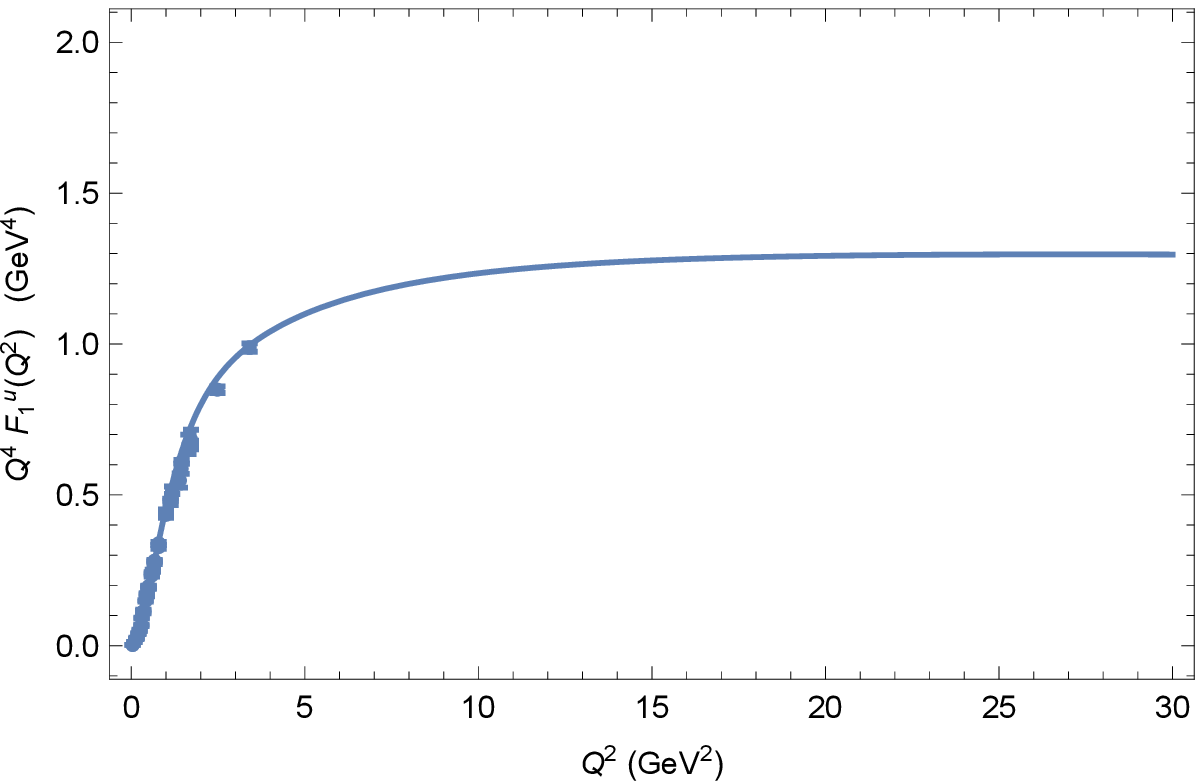,scale=.57}
\end{center}
\vspace*{-.5cm}
\noindent
\caption{Dirac $u$ quark
form factor multiplied by $Q^4$.
\label{fig:F1uQ4}}

\vspace*{.15cm}
\begin{center}
\epsfig{figure=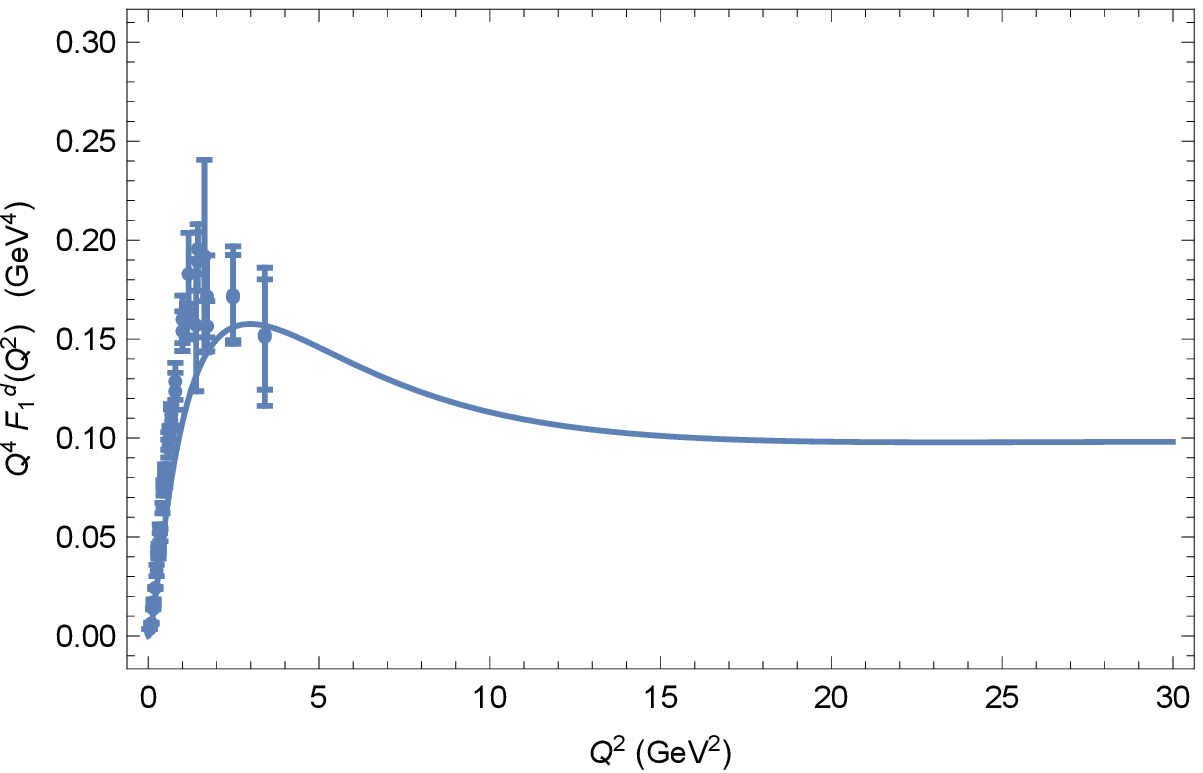,scale=.57}
\end{center}
\vspace*{-.5cm}
\noindent
\caption{Dirac $d$ quark
form factor multiplied by $Q^4$.
\label{fig:F1dQ4}}

\end{figure}

\begin{figure}

\begin{center}
\epsfig{figure=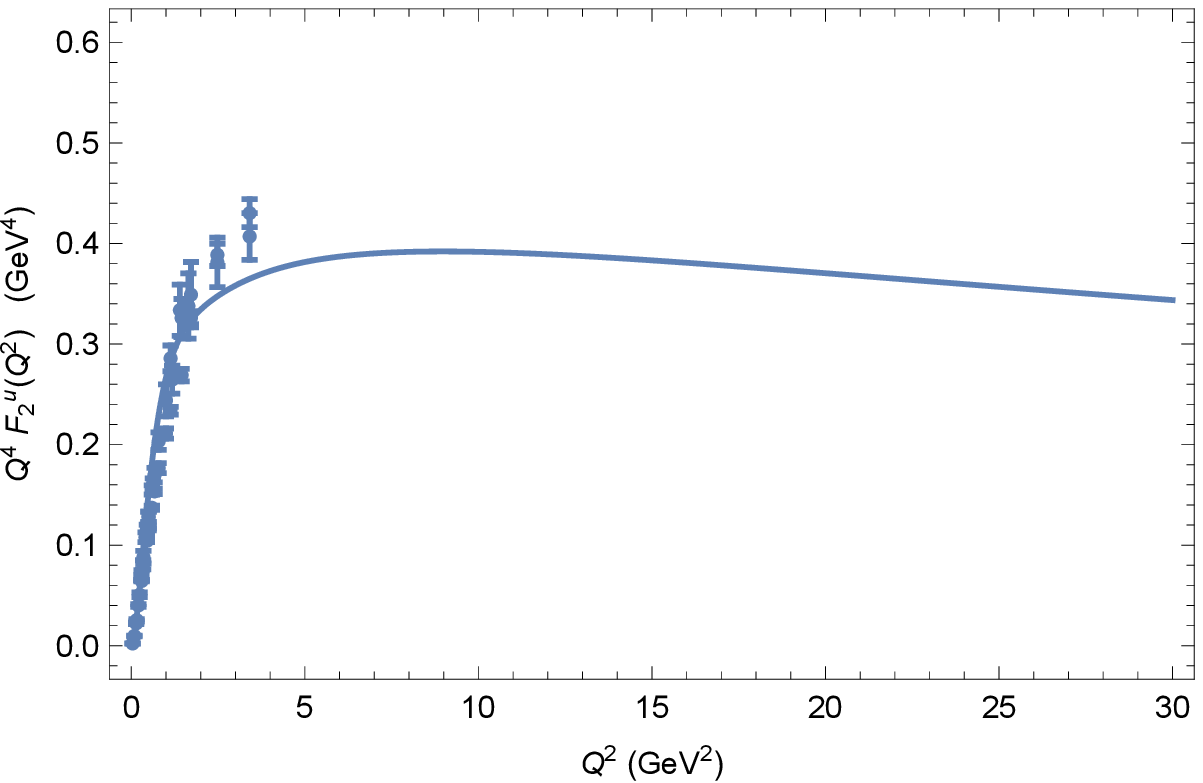,scale=.57}
\end{center}
\vspace*{-.5cm}
\noindent
\caption{Pauli $u$ quark
form factor multiplied by $Q^4$.
\label{fig:F2uQ4}}

\begin{center}
\epsfig{figure=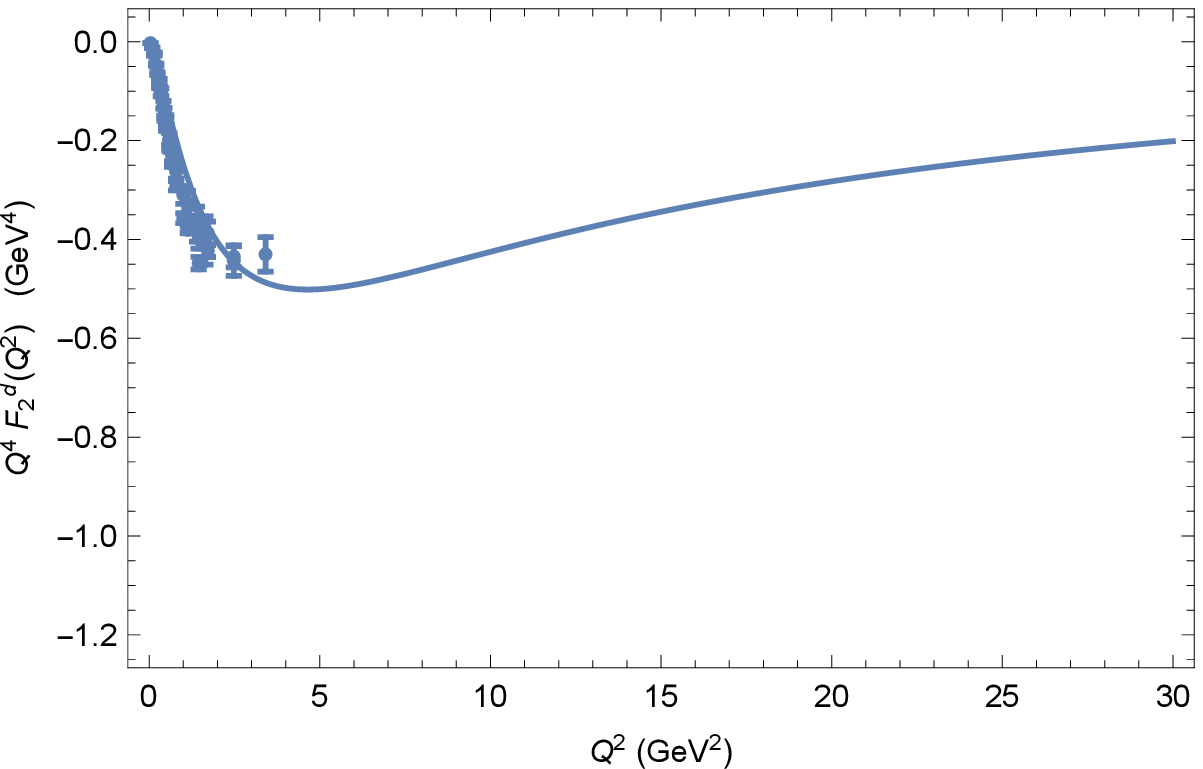,scale=.57}
\end{center}
\vspace*{-.5cm}
\noindent
\caption{Pauli $d$ quark
form factor multiplied by $Q^4$.
\label{fig:F2dQ4}}

\begin{center}
\epsfig{figure=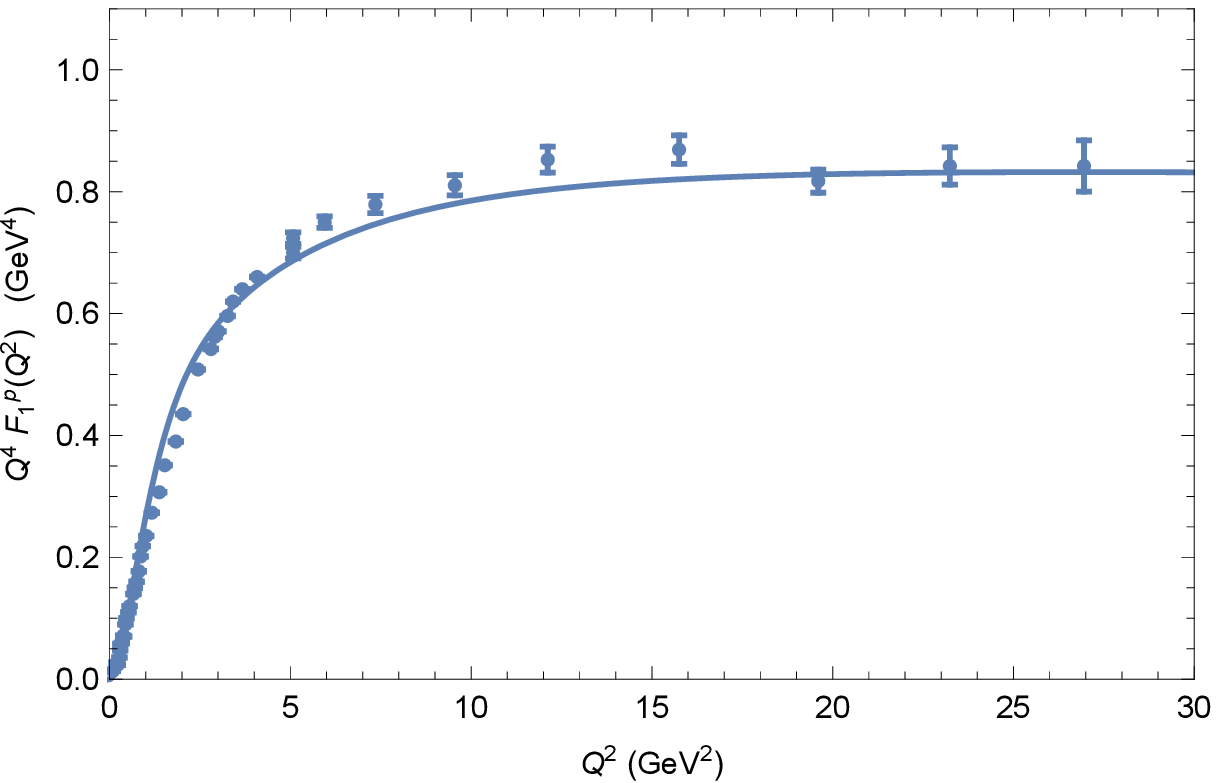,scale=.57}
\end{center}
\vspace*{-.5cm}
\noindent
\caption{Dirac proton
form factor multiplied by $Q^4$.
\label{fig:F1pQ4}}

\begin{center}
\epsfig{figure=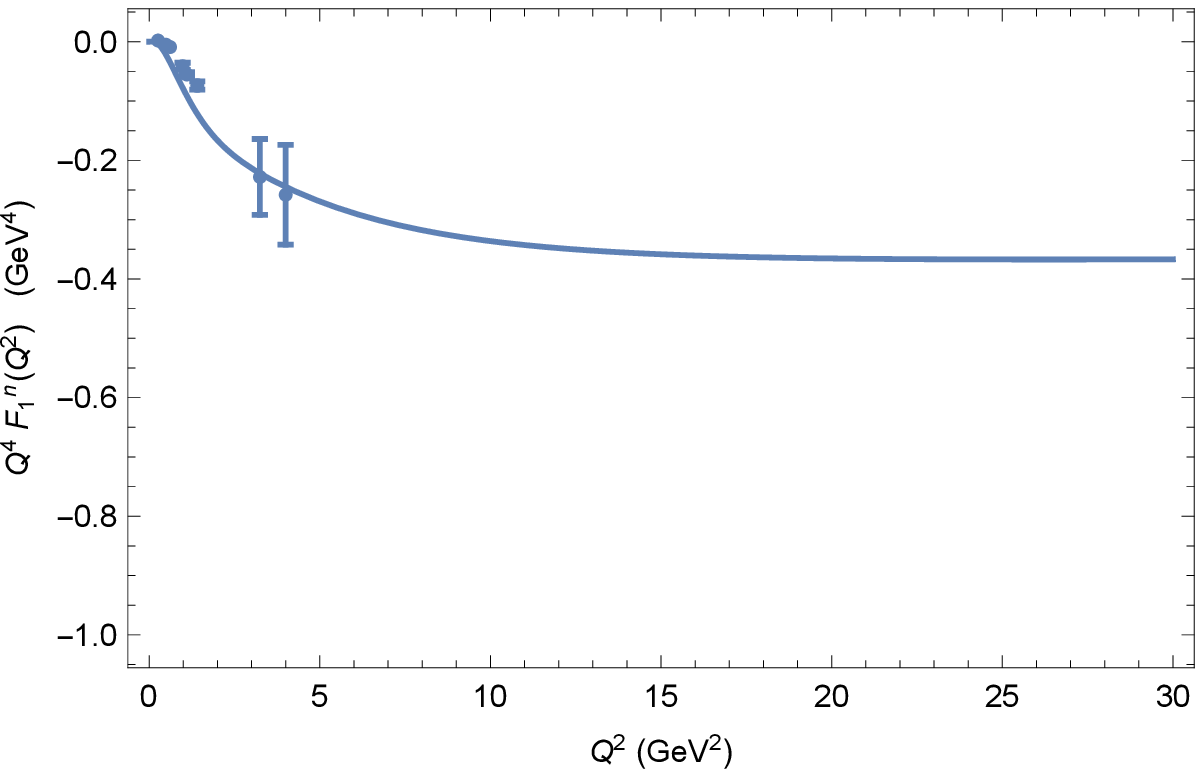,scale=.57}
\end{center}
\vspace*{-.5cm}
\noindent
\caption{Dirac neutron
form factor multiplied by $Q^4$.
\label{fig:F1nQ4}}

\end{figure}

\begin{figure}

\begin{center}
\epsfig{figure=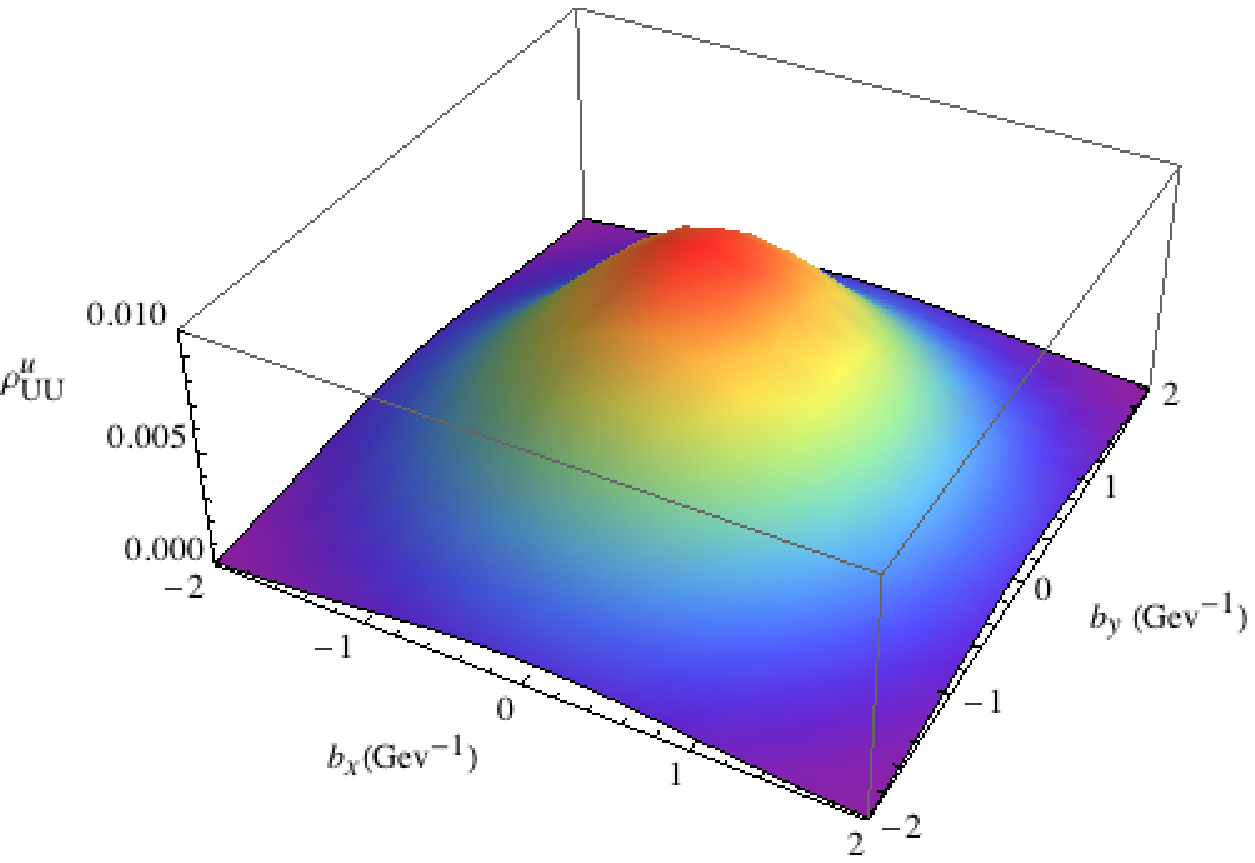,scale=.575}
\end{center}
\vspace*{-.8cm}
\noindent
\caption{Wigner distribution $\rho_{UU}^u(x,\bfb,\bfk)$ at $x=0.5$, \\
$k_x = k_y = 0.5$ GeV. 
\label{fig:rhoUUu}}

\vspace*{-.25cm}

\begin{center}
\epsfig{figure=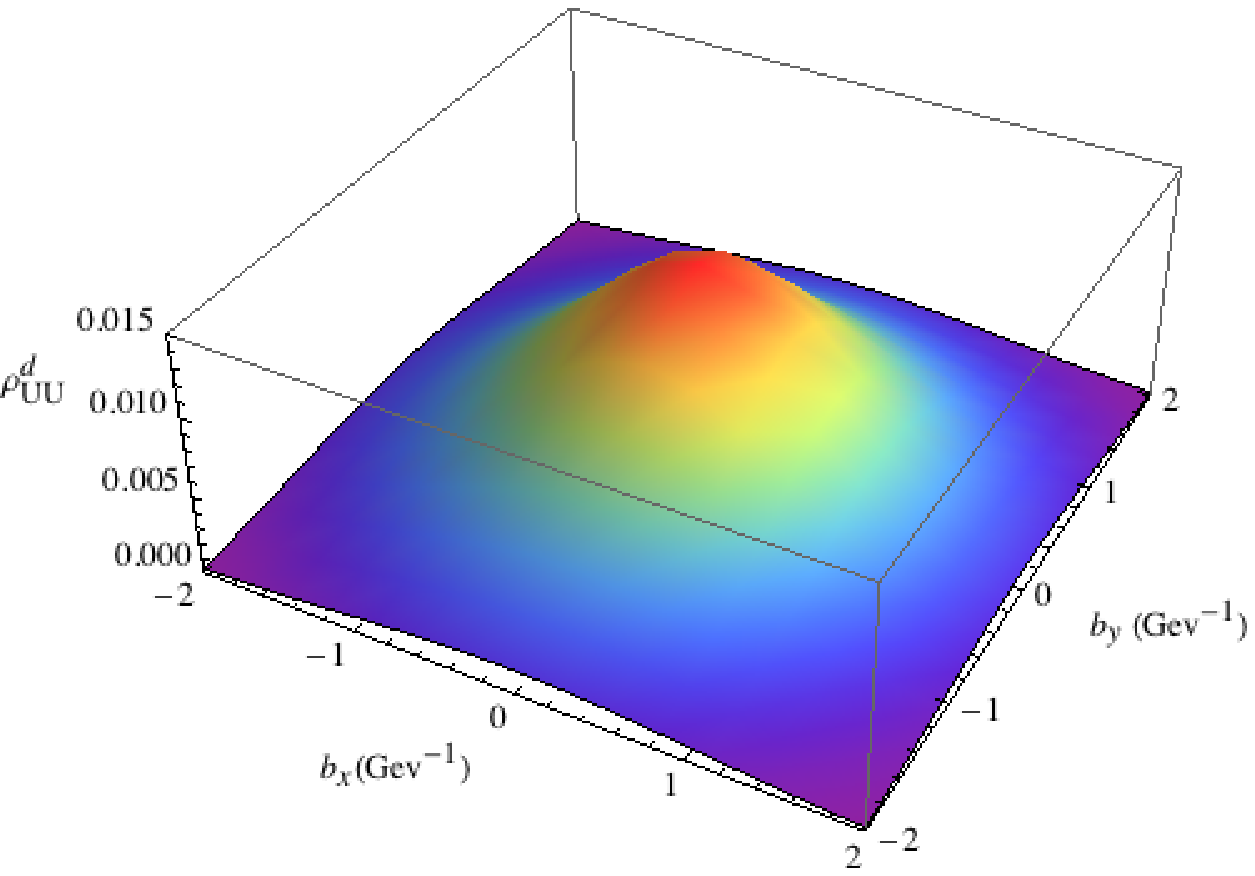,scale=.575}
\end{center}
\vspace*{-.8cm}
\noindent
\caption{Wigner distribution $\rho_{UU}^d(x,\bfb,\bfk)$ at $x=0.5$, \\
$k_x = k_y = 0.5$ GeV. 
\label{fig:rhoUUd}}

\vspace*{-.25cm}

\begin{center}
\epsfig{figure=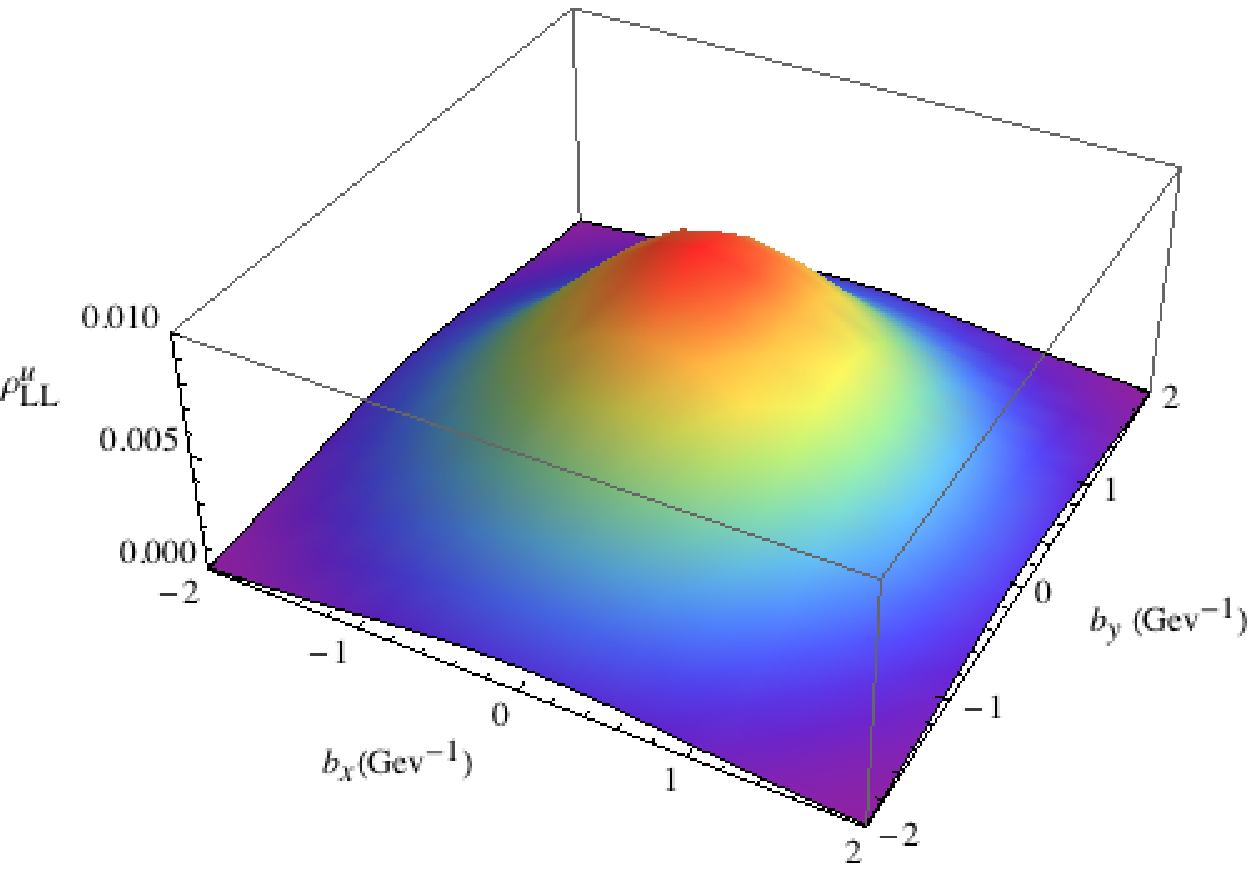,scale=.575}
\end{center}
\vspace*{-.8cm}
\noindent
\caption{Wigner distribution $\rho_{LL}^u(x,\bfb,\bfk)$ at $x=0.5$, \\
$k_x = k_y = 0.5$ GeV. 
\label{fig:rhoLLu}}

\vspace*{-.25cm}

\begin{center}
\epsfig{figure=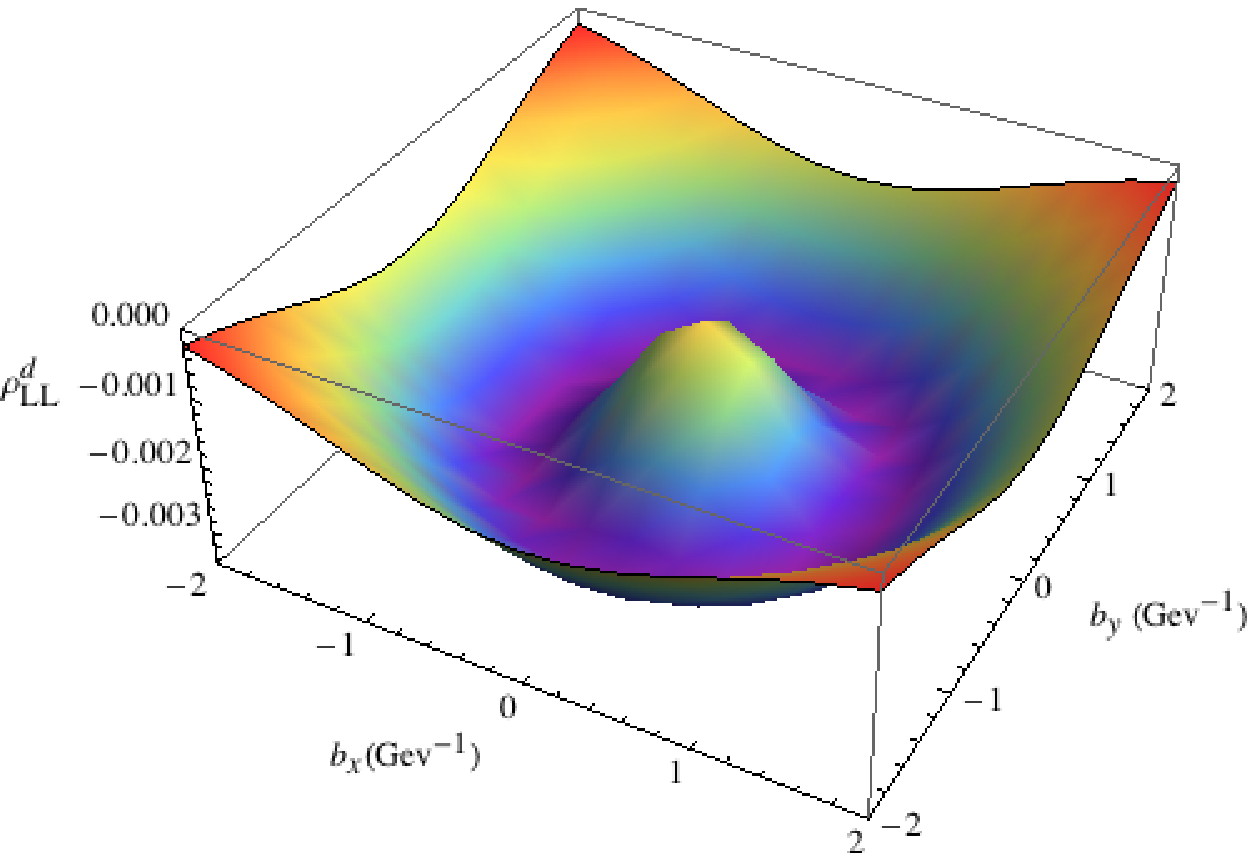,scale=.575}
\end{center}
\vspace*{-.8cm}
\noindent
\caption{Wigner distribution $\rho_{LL}^d(x,\bfb,\bfk)$ at $x=0.5$, \\
$k_x = k_y = 0.5$ GeV. 
\label{fig:rhoLLd}}

\end{figure}

\begin{figure}

\begin{center}
\epsfig{figure=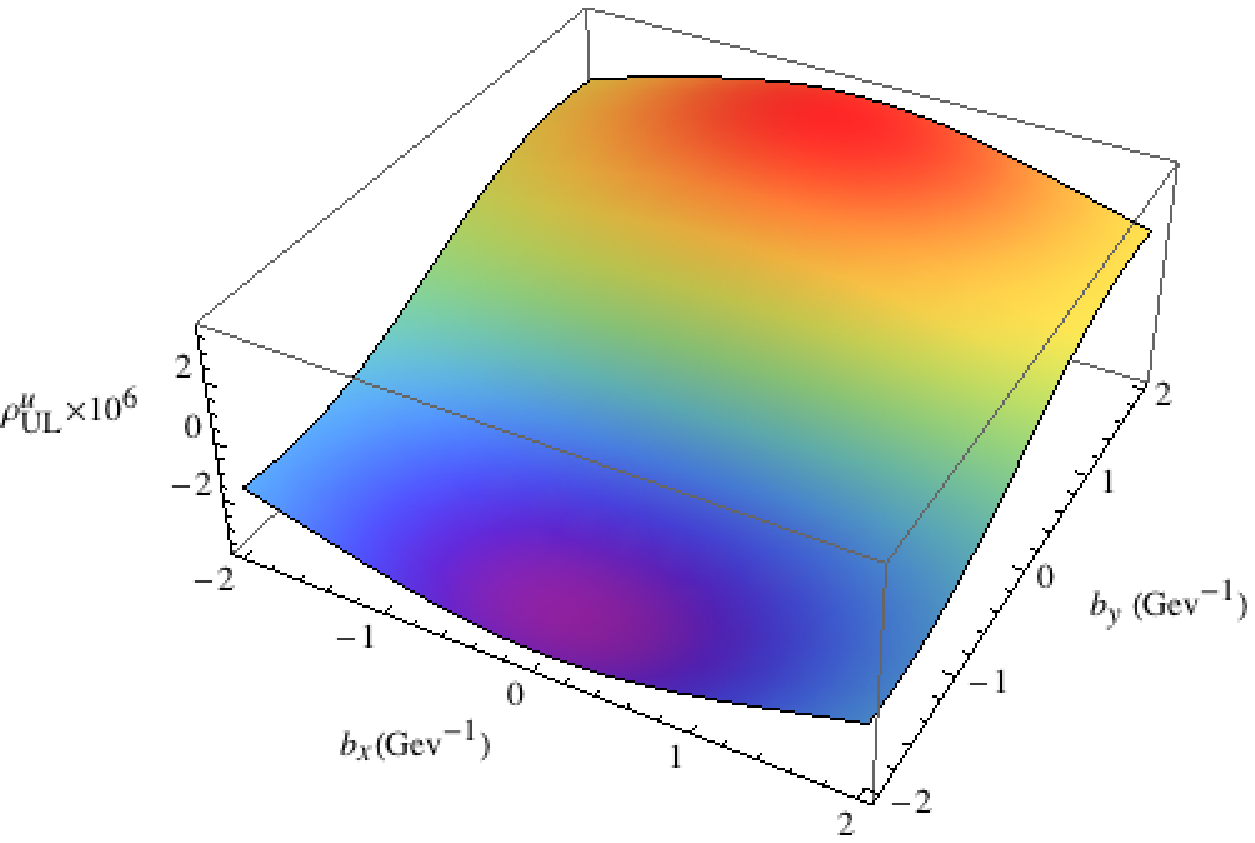,scale=.575}
\end{center}
\vspace*{-.5cm}
\noindent
\caption{Wigner distribution $\rho_{UL}^u(x,\bfb,\bfk)$ at $x=0.5$, \\
$k_x = k_y = 0.5$ GeV. 
\label{fig:rhoULu}}

\begin{center}
\epsfig{figure=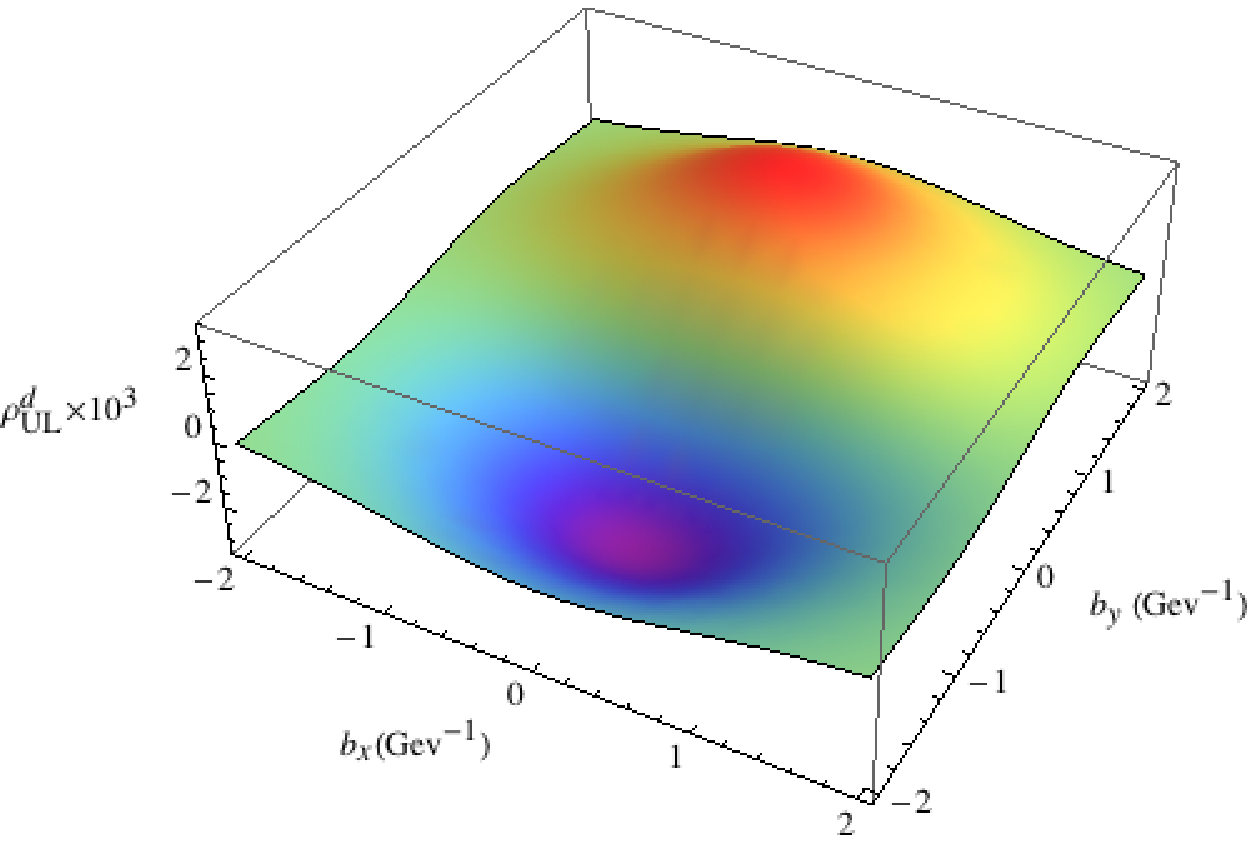,scale=.575}
\end{center}
\vspace*{-.5cm}
\noindent
\caption{Wigner distribution $\rho_{UL}^d(x,\bfb,\bfk)$ at $x=0.5$, \\
$k_x = k_y = 0.5$ GeV. 
\label{fig:rhoULd}}

\begin{center}
\epsfig{figure=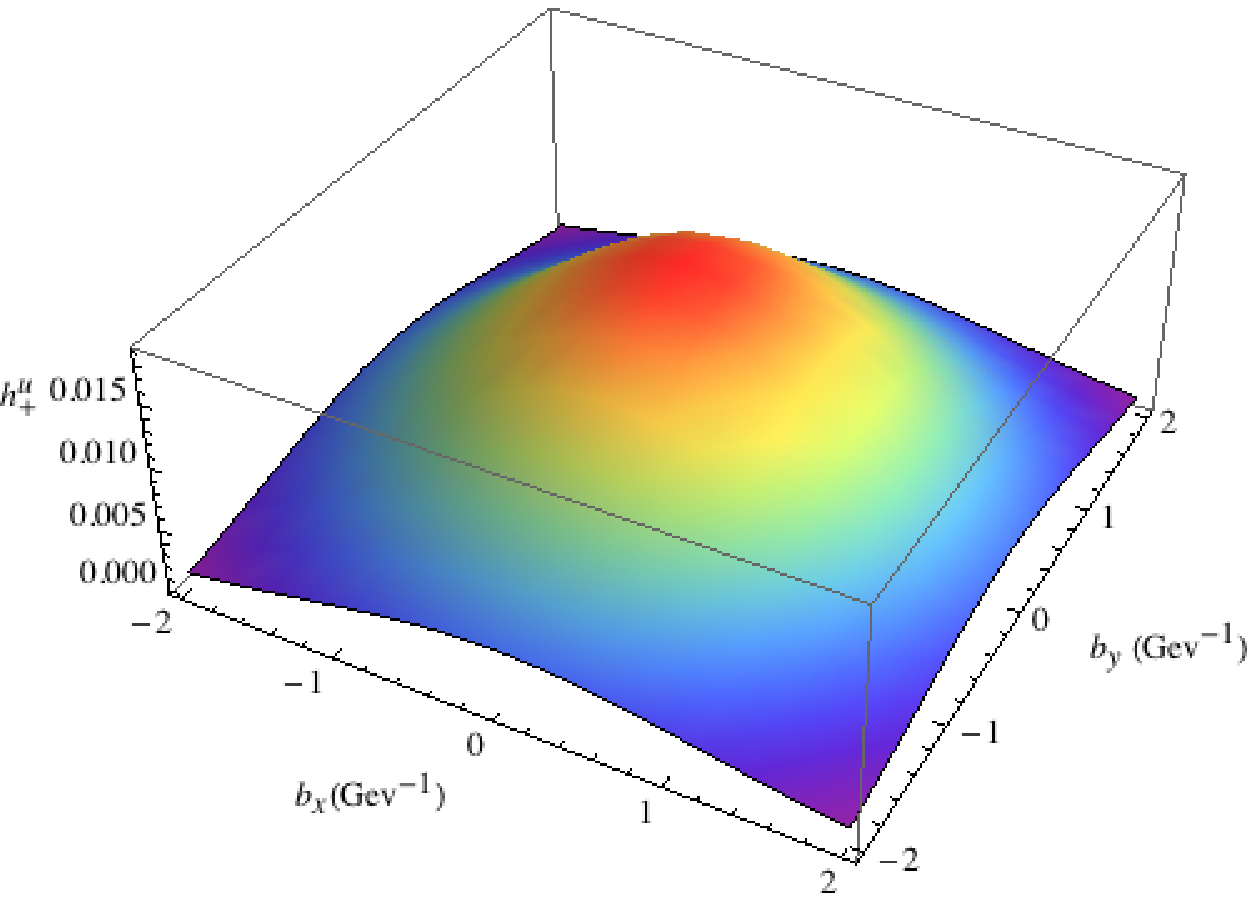,scale=.575}
\end{center}
\vspace*{-.5cm}
\noindent
\caption{Husimi distribution $h_+^u(x,\bfb,\bfk)$ at $x=0.5$, \\
$l = 1$ GeV$^{-1}$,  
$k_x = k_y = 0.5$ GeV. 
\label{fig:hpu}}

\end{figure}

\begin{figure}

\begin{center}
\epsfig{figure=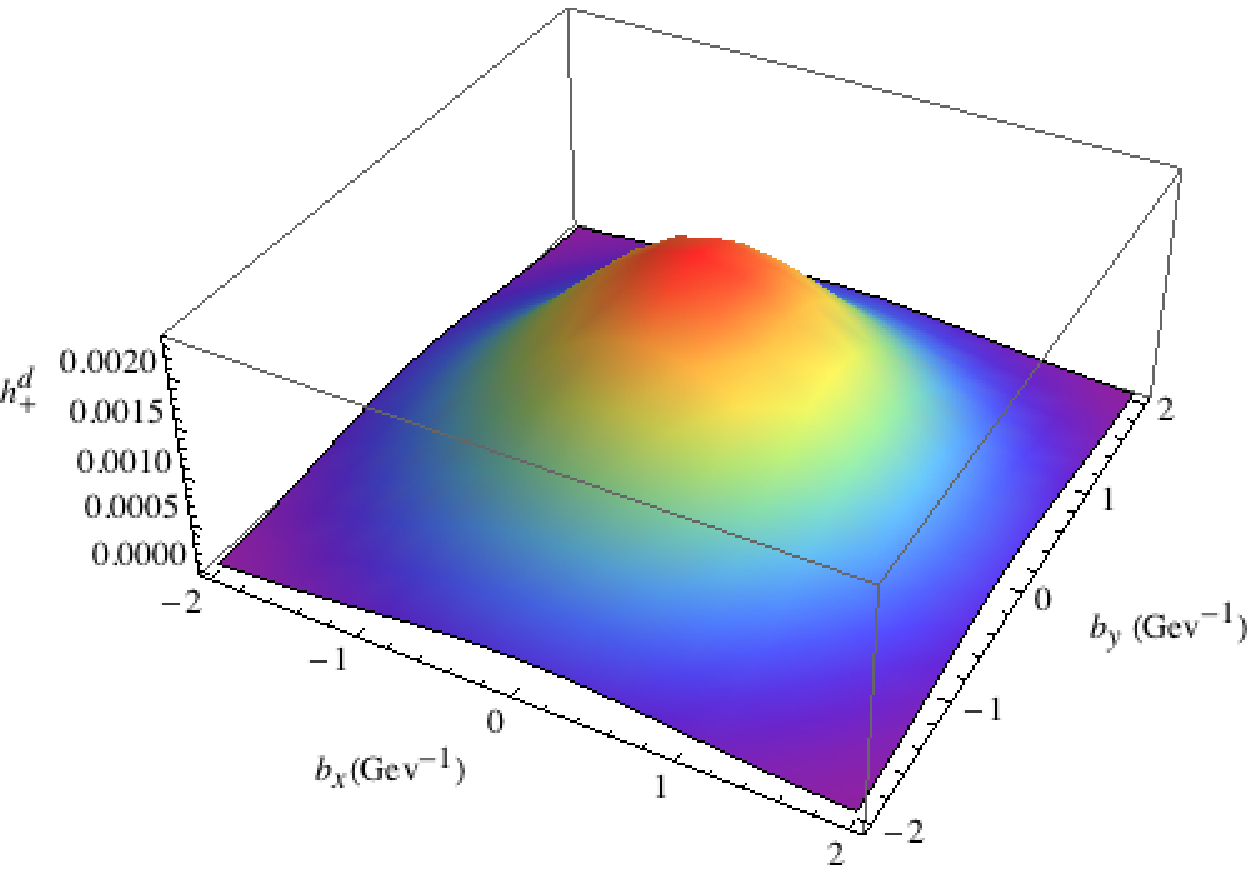,scale=.575}
\end{center}
\vspace*{-.5cm}
\noindent
\caption{Husimi distribution $h_+^d(x,\bfb,\bfk)$ at $x=0.5$, \\
$l = 1$ GeV$^{-1}$,  
$k_x = k_y = 0.5$ GeV. 
\label{fig:hpd}}

\begin{center}
\epsfig{figure=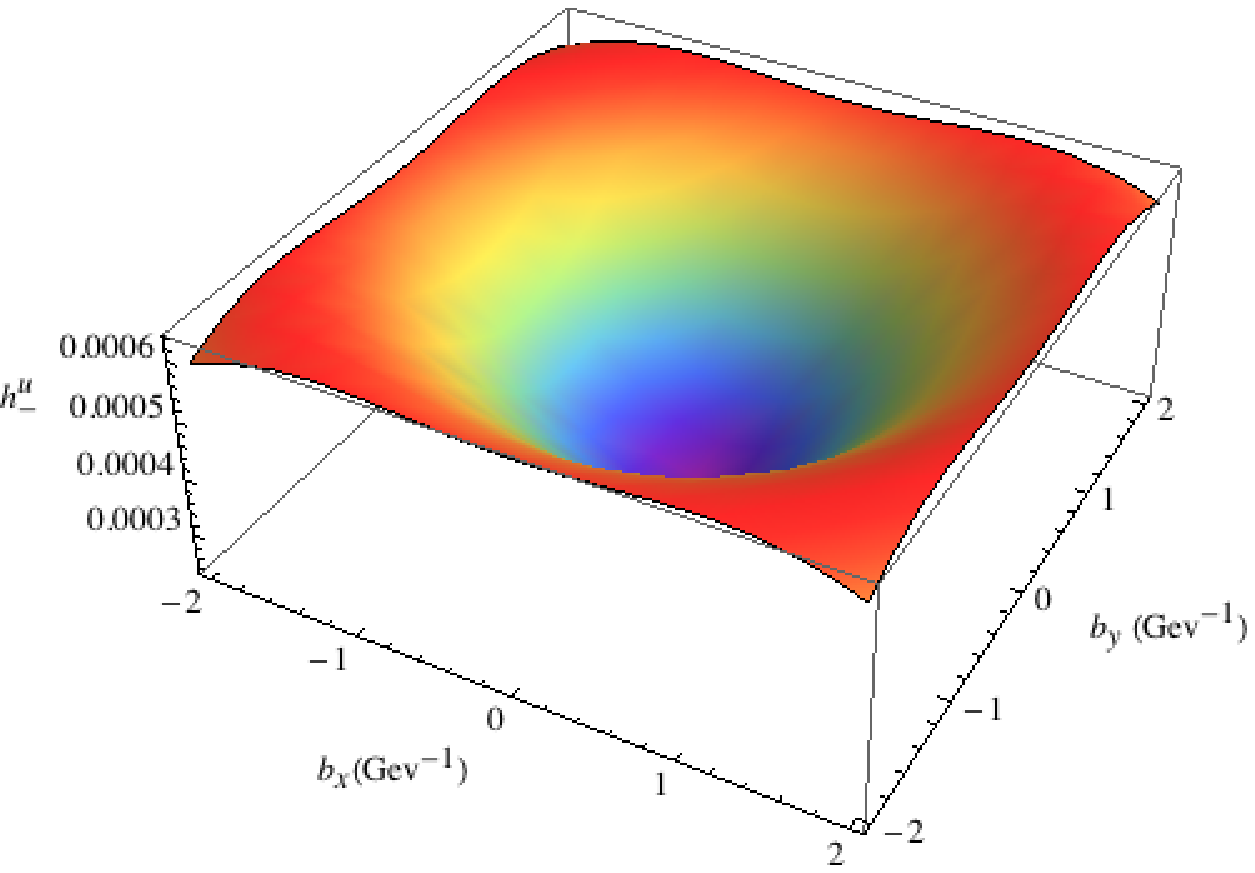,scale=.575}
\end{center}
\vspace*{-.5cm}
\noindent
\caption{Husimi distribution $h_-^u(x,\bfb,\bfk)$ at $x=0.5$, \\
$l = 1$ GeV$^{-1}$,  
$k_x = k_y = 0.5$ GeV. 
\label{fig:hmu}}

\begin{center}
\epsfig{figure=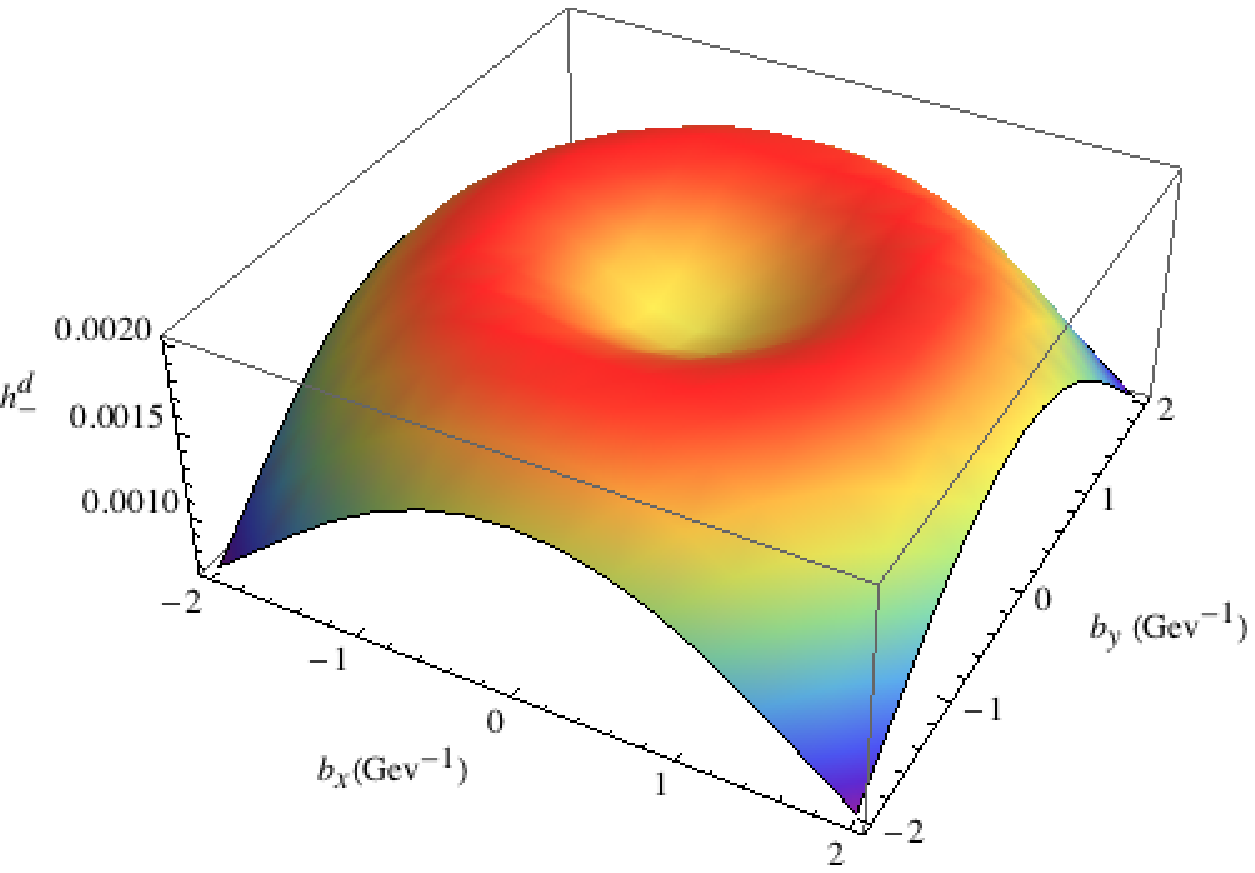,scale=.575}
\end{center}
\vspace*{-.5cm}
\noindent
\caption{Husimi distribution $h_-^d(x,\bfb,\bfk)$ at $x=0.5$, \\
$l = 1$ GeV$^{-1}$,  
$k_x = k_y = 0.5$ GeV. 
\label{fig:hmd}}
\end{figure}

In Figs.~\ref{fig:uPDF}-\ref{fig:hmd} 
we plot the results for the $x$-dependence of the unpolarized 
and polarized PDFs, TMDs, Wigner and Husimi distributions, 
and indicate selected results for the quark and 
nucleon electromagnetic form factors. The data on 
the quark decomposition of the nucleon form factors are taken 
from Refs.~\cite{Diehl:2013xca,Diehl:2006xca,Cates:2011pz}. 
In particular, in Fig.~\ref{fig:EqPDF} we show our predictions 
for magnetization PDFs ${\cal E}^q$ and compare them with results 
of Ref.~\cite{Guidal:2004nd}. 
In Fig.~\ref{fig:g1TLimits} we present 
a comparison of our predictions for 
$x g_{1T}^{q_v}(x)$ quark TMDs with 
the corresponding upper limits $x F_{1}^{q_v}(x)$. One can see that 
our results for $g_{1T}^{q_v}(x)$ are consistent with 
model-independent inequalities derived in Ref.~\cite{Bacchetta:1999kz}. 
Note that before in Sect.~\ref{Sec_LF_Decompositions} we proved it 
analytically. 
Our Wigner distributions are negative for longitudinal-logitudinal polarized
case of the $d$-quark and for unpolarized-longitudinal polarized case for
both quark flavors. Note that negative Wigner distributions 
have been obtained in some approaches, e.g. after including the gluons 
(see discussion 
in Refs.~\cite{Hagiwara:2014iya,Hatta:2015ggc,Mukherjee:2014nya}). 
 
\section{Conclusion} 
\label{Sec_Conclusion}

We want to summarize the main result of our paper. 
In the quark-scalar diquark picture we propose LFWFs for the nucleon 
which analytically reproduce the quark PDFs in the nucleon
at the initial scale $\mu \sim 1$ GeV. Our LFWFs contain four 
longitudinal wave functions $D_q^{(i)}$, $q=u, d$ and $i=1, 2$ 
depending on the $x$ variable, which 
are fixed from the analysis of nucleon form factors.  
Then we present a list of different types of nucleon parton distributions 
(TMDs, Wigner and Husimi distributions) in terms of the quark PDFs and the
longitudinal functions $D_q^{(i)}$. Finally, we present the numerical analysis 
for the quark distributions in the nucleon, we also indicate selected results 
for the quark and nucleon form factors using a specific ansatz for 
the NLO helicity-independent and helicity-dependent parton distributions at 
$\mu_{\rm NLO} = 0.40$ GeV$^2$~\cite{Gluck:1998xa, Gluck:2000dy}.
The resulting valence quark densities in the nucleon (e.g. TMDs) 
can be evolved to higher scales and can be compared 
to results for these quantities extracted in a data analysis. 

\vspace*{.5cm}
\begin{acknowledgments}

The authors thank Stan Brodsky, Lev Lipatov, 
Oleg Teryaev, Werner Vogelsang, 
and Marat Siddikov for useful discussions.  
This work was supported
by the German Bundesministerium f\"ur Bildung und Forschung (BMBF)
under Project 05P2015 - ALICE at High Rate (BMBF-FSP 202):
``Jet- and fragmentation processes at ALICE and the parton structure 
of nuclei and structure of heavy hadrons'',
by Tomsk State University Competitiveness
Improvement Program and the Russian Federation program ``Nauka''
(Contract No. 0.1526.2015, 3854),
by CONICYT (Chile) Research Project No. 80140097
and under Grants No. 7912010025, 1140390 and PIA/Basal FB0821. 
VEL would like to thank Departamento de F\'\i sica y Centro
Cient\'\i fico Tecnol\'ogico de Valpara\'\i so (CCTVal), Universidad 
T\'ecnica Federico Santa Mar\'\i a, Valpara\'\i so, Chile for warm hospitality.
In memory of my parents (VEL).

\end{acknowledgments}

\begin{widetext}
\appendix
\section{Wigner and Husimi parton distributions
 in the light-front quark model} 
\label{AppendixA}

The Wigner distributions 
$\rho^q_{UU}(\bfb,\bfk,x)$ and $\omega^q_{UU}(\bfD,\bfk,x)$ 
read 
\eq 
\rho^q_{UU} &=& \frac{1}{\pi^2 (1-x)^2} \, 
\biggl[ \frac{q_v(x)+\delta q_v(x)}{2} \, 
e^{-\frac{\bfk^2}{M_N^2} \, \alpha_q^{(1)}(x)} \, 
e^{-\bfb^2 \, M_N^2 \, \beta_q^{(1)}(x)}
\nonumber\\
&+& 
\frac{q_v(x)-\delta q_v(x)}{2} \, 
e^{-\frac{\bfk^2}{M_N^2} \, \alpha_q^{(2)}(x)} \, 
e^{-\bfb^2 \, M_N^2 \, \beta_q^{(2)}(x)} \, 
\biggl[ 
-1 \, + \, 
\frac{\bfk^2}{M_N^2} \, \alpha_q^{(2)}(x) 
\, + \, 
\bfb^2 \, M_N^2 \, \beta_q^{(2)}(x) \biggr] \biggr] 
\,, \\
\rho^q_{LL} &=& \frac{1}{\pi^2 (1-x)^2} \, 
\biggl[ \frac{q_v(x)+\delta q_v(x)}{2} \, 
e^{-\frac{\bfk^2}{M_N^2} \, \alpha_q^{(1)}(x)} \, 
e^{-\bfb^2 \, M_N^2 \, \beta_q^{(1)}(x)}
\nonumber\\
&-& 
\frac{q_v(x)-\delta q_v(x)}{2} \, 
e^{-\frac{\bfk^2}{M_N^2} \, \alpha_q^{(2)}(x)} \, 
e^{-\bfb^2 \, M_N^2 \, \beta_q^{(2)}(x)} \, 
\biggl[ 
-1 \, + \, 
\frac{\bfk^2}{M_N^2} \, \alpha_q^{(2)}(x) 
\, + \, 
\bfb^2 \, M_N^2 \, \beta_q^{(2)}(x) \biggr] \biggr] 
\,, \\
\rho^q_{UL} &=& - \rho^q_{LU} \, = \, 
\frac{1}{\pi^2 (1-x)^3} \, \epsilon^{ij} \, \bfk^i \, \bfb^j \, 
\Big[q_v(x)-\delta q_v(x)\Big] \, 
e^{-\frac{\bfk^2}{M_N^2} \, \alpha_q^{(2)}(x)} \, 
e^{-\bfb^2 \, M_N^2 \, \beta_q^{(2)}(x)} 
\en 
and 
\eq 
\omega^q_{UU}(\bfD,\bfk,x) &=& \frac{1}{\pi M_N^2} \, 
\biggl[ \frac{q_v(x)+\delta q_v(x)}{2} \, \alpha_q^{(1)}(x) \, 
e^{-\frac{\bfk^2+ \frac{\bfD^2}{4} (1-x)^2}{M_N^2} \, \alpha_q^{(1)}(x)} \, 
\nonumber\\
&+& 
\frac{q_v(x)-\delta q_v(x)}{2} \, \Big(\alpha_q^{(2)}(x)\Big)^2 \, 
\frac{\bfk^2 + \frac{\bfD^2}{4} (1-x)^2}{M_N^2} \, 
e^{-\frac{\bfk^2 + \frac{\bfD^2}{4} (1-x)^2}{M_N^2} \, \alpha_q^{(1)}(x)} 
\biggr]  \,, \\
\omega^q_{LL}(\bfD,\bfk,x) &=& \frac{1}{\pi M_N^2} \, 
\biggl[ \frac{q_v(x)+\delta q_v(x)}{2} \, \alpha_q^{(2)}(x) \, 
e^{-\frac{\bfk^2+ \frac{\bfD^2}{4} (1-x)^2}{M_N^2} \, \alpha_q^{(2)}(x)} \, 
\nonumber\\
&-& 
\frac{q_v(x)-\delta q_v(x)}{2} \, \Big(\alpha_q^{(2)}(x)\Big)^2 \, 
\frac{\bfk^2 + \frac{\bfD^2}{4} (1-x)^2}{M_N^2} \, 
e^{-\frac{\bfk^2 + \frac{\bfD^2}{4} (1-x)^2}{M_N^2} \, \alpha_q^{(2)}(x)} 
\biggr]  \,, \\ 
\omega^q_{UL}(\bfD,\bfk,x) &=& - \omega^q_{LU}(\bfD,\bfk,x)  \, = \, 
\frac{1}{\pi M_N^4} \, i \, \epsilon^{ij} \, \bfk^i \, \bfD^j \, 
\Big[q_v(x)-\delta q_v(x)\Big] \, 
e^{-\frac{\bfk^2}{M_N^2} \, \alpha_q^{(2)}(x)} \, 
e^{-\bfb^2 \, M_N^2 \, \beta_q^{(2)}(x)} \,, 
\en 
where $\alpha_q^{(i)}(x) = D_q^{(i)}(x)$, 
$\beta_q^{(i)}(x) = \frac{1}{(1-x)^2 \, D_q^{(i)}(x)}$, 
$\epsilon^{12} = - \epsilon^{21} = 1$.  

The integrals over the Wigner distributions are related to the 
TMDs, NPDs and PDFs by  
\eq 
& &\int d^2\bfb \rho^q_{UU}(\bfb,\bfk,x) = f_1^q(x,\bfk) \,, \nonumber\\
& &\int d^2\bfb \rho^q_{LL}(\bfb,\bfk,x) = g_{1L}^q(x,\bfk)\,,  
\en 
\eq 
& &\int d^2\bfb \omega^q_{UU}(\bfD,\bfk,x) = {\cal H}(x,0,\bfD^2) 
\,, \nonumber\\
& &\int d^2\bfb \omega^q_{LL}(\bfD,\bfk,x) = \tilde{\cal H}(x,0,\bfD^2) \,, 
\en 
and 
\eq 
& &\int d^2\bfk d^2\bfb \rho^q_{UU}(x,\bfk,\bfb) = q_v(x) = f_1^{q_v}(x) 
\,, \nonumber\\
& &\int d^2\bfk d^2\bfb \rho^q_{LL}(x,\bfk,\bfb) = \delta q_v(x) = g_{1L}^{q_v}(x)
\en 
and 
\eq 
& &\int\limits_0^1 dx 
\int d^2\bfk d^2\bfb \rho^q_{UU}(x,\bfk,\bfb) = n_q 
\,, \nonumber\\
& &\int\limits_0^1 dx 
\int d^2\bfk d^2\bfb \rho^q_{LL}(x,\bfk,\bfb) = g_A^q
\,, \nonumber\\
& &\int\limits_0^1 dx 
\int d^2\bfk d^2\bfb \rho^q_{UL}(x,\bfk,\bfb) = 
\int\limits_0^1 dx 
\int d^2\bfk d^2\bfb \rho^q_{LU}(x,\bfk,\bfb) = 0 \,. 
\en 

The Husimi parton distributions are given by 
\eq 
h_{UU}^q(x,\bfb,\bfk) + h_{LL}^q(x,\bfb,\bfk) 
&=& \frac{M_N^2 l^2}{\pi^2} \, 
(q_v(x)+\delta q_v(x)) 
\, \rho_q^{(1)}(x) \, \sigma_q^{(1)}(x)  \, 
e^{-\bfk^2 \, l^2 \, \rho_q^{(1)}(x)} \, 
e^{-\bfb^2 \, M_N^2 \, \sigma_q^{(1)}(x)} \,, \\
h_{UU}^q(x,\bfb,\bfk) - h_{LL}^q(x,\bfb,\bfk) 
&=& \frac{M_N^2 l^2}{\pi^2} \, 
(q_v(x)-\delta q_v(x)) 
\, \rho_q^{(2)}(x) \, \sigma_q^{(2)}(x)  \, 
e^{-\bfk^2 \, l^2 \, \rho_q^{(2)}(x)} \, 
e^{-\bfb^2 \, M_N^2 \, \sigma_q^{(2)}(x)} \nonumber\\
&\times& 
\biggl[ 1 + \frac{M_N^2 l^2}{D_q^{(2)}(x)} \, \rho_q^{(2)}(x) 
\Big(\bfk^2 l^2 \rho_q^{(2)}(x) - 1 \Big)\nonumber\\
&+& 
D_q^{(2)}(x) \, \sigma_q^{(2)}(x) (1-x)^2 \, 
\Big(\bfb^2 M_N^2 \sigma_q^{(2)}(x) - 1 \Big)\biggr] \,, \\  
h_{UL}^q(x,\bfb,\bfk) &=& - h_{LU}^q(\bfb,\bfk,x) \nonumber\\
&=& 
\frac{1}{\pi^2} \, \epsilon^{ij} \, \bfk^i \, \bfb^j \, 
\Big[q_v(x)-\delta q_v(x)\Big] \, (1-x) \, M_N^4 l^4 \, 
\biggl[\rho_q^{(2)}(x) \, \sigma_q^{(2)}(x)\biggr]^2 \nonumber\\
&\times&
e^{-\bfk^2 \, l^2 \, \rho_q^{(2)}(x)} \, 
e^{-\bfb^2 \, M_N^2 \, \sigma_q^{(2)}(x)} \,, 
\en   
where 
\eq 
\rho_q^{(i)}(x) = \frac{D_q^{(i)}(x)}{M_N^2 l^2 + D_q^{(i)}(x)} \,, 
\quad 
\sigma_q^{(i)}(x) = \frac{1}{M_N^2 l^2 + D_q^{(i)}(x) (1-x)^2} \,. 
\en 
The expressions for the entropies of the nucleon $S^q_\pm(x)$ are given by 
\eq 
S^q_+(x) &=& (q_v(x)+\delta q_v(x)) 
\, \biggl[1 - \frac{1}{2} \log\Big(\frac{q_v(x)+\delta q_v(x)}{2\pi^2}\Big)  
\biggr] \,, \nonumber\\
S^q_-(x) &=& (q_v(x)-\delta q_v(x)) 
\, \biggl[1 - \frac{1}{2} \log\Big(\frac{q_v(x)-\delta q_v(x)}{2\pi^2}\Big)  
\nonumber\\
&-& \frac{1}{2} \log(B) - \frac{A_1+A_2}{4}
\biggl( A_1 \int\limits_0^\infty \frac{dt e^{-t}}{A_1t+B}
      + A_2 \int\limits_0^\infty \frac{dt e^{-t}}{A_2t+B}
\biggr) \nonumber\\
&-& \frac{A_1^2+A_2^2}{4} \, 
\int\limits_0^\infty\int\limits_0^\infty 
\frac{dt_1 dt_2 e^{-t_1-t_2}}{A_1t_1+A_2t_2+B}
\biggr] 
\en 
where 
\eq 
A_1 = \frac{D_q^{(2)}(x) (1-x)^2}{M_N^2 l^2 + D_q^{(2)}(x) (1-x)^2} 
\,, \ 
A_2 = \frac{M_N^2 l^2}{M_N^2 l^2 + D_q^{(2)}(x)} \,, \ 
B = 1 - A_1 - A_2 \,. 
\en 

\end{widetext}

\end{document}